\documentclass[apj]{emulateapj}
\usepackage[colorlinks,linkcolor={blue},citecolor={blue},urlcolor={red}]{hyperref}
\usepackage{bm}
\usepackage{graphicx}
\usepackage{epsf}
\usepackage{graphics}
\usepackage{amsmath}
\usepackage{lineno}

\def\beq{\begin{equation}}
\def\eeq{\end{equation}}
\def\bey{\begin{eqnarray}}
\def\eey{\end{eqnarray}}

\def\kms{\,{\rm {km\, s^{-1}}}}
\def\kpc{\,{\rm {kpc}}}
\def\Msun{{\rm M_\odot}}

\def\gs{\mathrel{\raise1.16pt\hbox{$>$}\kern-7.0pt
\lower3.06pt\hbox{{$\scriptstyle \sim$}}}}
\def\ls{\mathrel{\raise1.16pt\hbox{$<$}\kern-7.0pt
\lower3.06pt\hbox{{$\scriptstyle \sim$}}}}
\def\gtsima{\, {\buildrel > \over \sim} \,}
\def\ltsima{\, {\buildrel < \over \sim} \,}
\def\prosima{\, {\buildrel \propto \over \sim} \,}
\def\gsim{\lower.5ex\hbox{\gtsima}}
\def\lsim{\lower.5ex\hbox{\ltsima}}
\def\simgt{\lower.5ex\hbox{\gtsima}}
\def\simlt{\lower.5ex\hbox{\ltsima}}
\def\simpr{\lower.5ex\hbox{\prosima}}

\shorttitle{Hotness, quenching and SMBH}
\shortauthors{H. Hong et al.}

\begin{document}
\title{Dynamical hotness, star formation quenching and growth of supermassive black holes}

\author{Hui Hong\altaffilmark{1,2}, Huiyuan Wang\altaffilmark{1,2}, H. J. Mo\altaffilmark{3}, Ziwen Zhang\altaffilmark{1,2}, Guangwen Chen\altaffilmark{1,2}, Wentao Luo\altaffilmark{1,2}, Tinggui Wang\altaffilmark{1,2}, Pengfei Li\altaffilmark{4}, Renjie Li\altaffilmark{1,2}, Yao Yao\altaffilmark{1,2}, and Aoxiang Jiang\altaffilmark{1,2}}
\altaffiltext{1}{Key Laboratory for Research in Galaxies and Cosmology, Department of Astronomy, University of Science and Technology of China, Hefei, Anhui 230026, China;js2011@mail.ustc.edu.cn,whywang@ustc.edu.cn}
\altaffiltext{2}{School of Astronomy and Space Science, University of Science and Technology of China, Hefei 230026, China}
\altaffiltext{3}{Department of Astronomy, University of Massachusetts, Amherst MA 01003-9305, USA}
\altaffiltext{4}{Department of Physics and Astronomy, University of Utah, UT 84112, USA}

\begin{abstract}
A stellar system is dynamically hot when its kinetic energy is dominated by random motion represented by the velocity dispersion $\sigma_{hot}$.  We use MaNGA data to obtain inner and outer dispersion of a galaxy,  $\sigma_{in}$ and $\sigma_{out}$, to characterize its dynamical status and study its connection with star-formation quenching and the growth of supermassive black hole(SMBH). We divide galaxies into fully quenched (FQGs), partially quenched (PQGs) and fully star-forming (FSGs) populations, and identify quenched central cores (QCCs) in PQGs. The galaxy distribution in $(\sigma_{in}/\sigma_{hot})$-$(\sigma_{out}/\sigma_{hot})$ diagram is L-shaped, consisting of a horizontal sequence ($\sigma_{out}/\sigma_{hot}\sim0$) and a vertical sequence ($\sigma_{in}/\sigma_{hot}\sim1$). FQGs and QCCs are located at the top of the vertical sequence,  $\sigma_{out}/\sigma_{hot}\sim1$, and are thus dynamically hot over their entire bodies. PQGs reside along the vertical sequence, so they have hot center but cold outskirt. FSGs are diverse and can be found in both sequences. Galaxy structural properties, star formation and AGN activities make a transition along the horizontal sequence at $\sigma_{in}/\sigma_{hot}\sim 0.5$, and along the vertical sequence at $\sigma_{out}/\sigma_{hot}\sim 0.5$. The fractions of optical AGNs and barred galaxies increase rapidly in the first transition and decline rapidly in the second; radio galaxies are located at the top of the vertical sequence. Our results demonstrate that star formation quenching and SMBH growth are effective only in dynamically hot systems. A simple model along this line can reproduce the observed SMBH scaling relations. We discuss how secular processes and strong interactions can make a system dynamically hot, and lead to the SMBH growth and 
star-formation quenching.
\end{abstract}


\section{Introduction}
\label{sec_intro}

The quenching of star formation in galaxies is one of the key processes 
during the evolution of the galaxy population. Observations show that   
star formation quenching began quite early, at least at redshift $z=3$ 
\citep{Muzzin2013,Straatman2014,Ilbert2013}, and generated a population of 
galaxies that are distinct from the star-forming population in 
color, morphology, gas content, and large-scale clustering 
\citep[e.g.][]{Strateva2001, Baldry2004, Brinchmann2004, Zehavi2011, Fabello2011, WangH2018}. 
Thus, quenching seems to have played a key role in producing the rich diversity 
of the observed galaxy population. Theoretically, numerous hypotheses
have been proposed to understand the underlying mechanisms for 
star formation quenching \citep[see reviews in][]{Mo2010book, Fabian2012, Heckman2014, 
Wechsler2018, Bosseli2022}.

Many internal properties of galaxies, such as stellar mass, morphology, 
bulge mass, central stellar surface density, and central velocity dispersion, 
are found to correlate with the quenching properties of galaxies
\citep[][]{Baldry2004, Peng2010, Cheung2012, Fang2013ApJ, Bluck2014, 
Bluck2016, Woo2015, WangE2018b,ChenGuangwen2020ApJ,Bluck2022}.  For example, quiescent 
(quenched) galaxies tend to be more massive and compact than star-forming 
galaxies, have denser stellar cores and higher central velocity dispersion. 
Such correlations are particularly prominent for central galaxies, 
which are the most massive ones in dark matter halos, presumably 
because central galaxies are less affected by environmental effects
\citep[e.g.][]{LiP2020}. 
Recently, based on weak lensing and HI rotational curve data, it is revealed that massive disk galaxies have a very high baryon-to-star conversion efficiency \citep{Posti2019,ZhangZ2022}. It hints that star formation can not be quenched if morphological transformation does not happen. All of these suggest a connection between quenching and morphological transformation.

The internal properties of galaxies are correlated with each other. 
For example, the bulge-to-total mass ratio tends to increase with galaxy stellar mass, $M_*$, and the 
central stellar mass surface density is tightly correlated with the central 
velocity dispersion ($\sigma_{\rm in}$) 
\citep[e.g.][]{Fang2013ApJ, Saglia2016ApJ, Bluck2020}.
Among them, the well-known relation is the Faber-Jackson relation 
for elliptical and early-type galaxies, a tight relation 
between stellar mass and central velocity dispersion: $M_*\propto\sigma_{\rm in}^{3\sim4}$\citep{Faber1976, Gallazzi2006, Aquino2018, Oh2020}.
Thus, the observed correlation of quenching with internal properties 
alone does not provide causality between them.  

One of the most studied parameters is the central stellar mass surface 
density within $1 \kpc$, $\Sigma_1$ 
\citep[e.g.][]{Franx2008, Cheung2012, Fang2013ApJ, Barro2017ApJ, Ellison2018, Bluck2020}. 
Previous studies found that the fraction of quenched galaxies 
increases with $\Sigma_1$, particularly rapidly around 
$\log\Sigma_1/(\Msun/\kpc^2)\sim9.5$ \citep[e.g.][]{Cheung2012}.
Alone the same lines, \cite{Fang2013ApJ} found an L-shaped 
distribution of galaxies in $\Sigma_1$-color space, indicating 
a rapid transition from the star-forming (blue) to 
quiescent (red) populations at a characteristic value of $\Sigma_1$. 
\cite{Barro2017ApJ} extended the analysis to high redshift and found 
similar results. However, the central stellar mass surface density 
does not change monotonously with stellar mass for massive elliptical 
galaxies \citep[e.g.][]{Graham2003}. 
Moreover, \cite{Fang2013ApJ} found that 
quenched and star-forming galaxies have similar median stellar mass 
density profiles at large radii, even though the two populations  
have different morphology.   It suggests that the quenching is not regulated 
by stellar mass surface density alone.

The stellar velocity  dispersion, $\sigma$, is another quantity 
widely used to indicate the dynamic structure of a galaxy. For instance, dynamically 
cold stellar components, such as galaxy disks, usually have much lower 
velocity dispersion than dynamically hot components, such as bulges and 
elliptical galaxies \citep[e.g.][]{Brook2012, Cappellari2016, Zhu2018, Du2020, Zhu2022b}. 
\cite{Fang2013ApJ} found an L-shaped transition in the 
velocity dispersion-color diagram, similar to that seen 
in the $\Sigma_1$-color diagram. \cite{Bluck2020, Bluck2022} used a machine learning technique to analyze various correlations 
of quenching with other galaxy properties, and found that the 
central velocity dispersion, $\sigma_{\rm in}$, is the most predictive 
parameter for star formation quenching \citep[see also][]{Wake2012}. Furthermore, \cite{Brownson2022} found that the velocity dispersion 
performs better than other rotation-related dynamical parameters. Clearly, stellar velocity dispersion merits further investigation as a 
potentially key factor in driving galaxy quenching.  

The mass of supermassive black hole (SMBH) in 
a galaxy is found to be tightly correlated with the velocity dispersion 
and mass of the spheroid component of the host galaxy 
\citep[][]{Magorrian1998, Ferrarese2000, Gebhardt2000, Aller2007, Graham2008ApJ, 
Kormendy2013ARA&A, Saglia2016ApJ, Graham2023}. 
These correlations are usually referred to as the SMBH scaling relations (BHSRs).
The correlation of quenching properties with $\sigma_{\rm in}$ and $\Sigma_1$ 
may thus be a consequence of feedback from active galactic nuclei (AGNs).
Observationally, attempts have been made to find the best predictor of the  
SMBH mass($M_{\rm BH}$) based on various combinations of velocity 
dispersion and spheroid mass, $M_{\rm sph}$ \citep{Feoli2005, Aller2007, Hopkins2007b, Soker2011, Mancini2012}. 
More recently, \cite{Graham2023} found that 
the $M_{\rm BH}$-$M_{\rm sph}$ scaling relation for elliptical galaxies 
has a significant offset relative to that for bulges in other early-type and spiral  
galaxies \citep[see also][]{Bogdan2018ApJ, Sahu2019ApJ}.
This hints that the BHSRs in bulges and elliptical galaxies may 
not share the same origin. However, caution needs to be exercised 
before drawing such a conclusion. Indeed, simulations and observational 
data show that, in addition to bulges, galaxies contain other 
dynamically hot components
such as inner stellar halos discovered in nearby galaxies
\citep[see][]{Brook2012, Ibata2014,Du2020, Zhu2022a, Zhu2022b}, 
which may also be related to the formation of SMBHs.   
Clearly, more investigations are needed, in particular 
to understand the BHSRs from the perspective of galaxy 
quenching. 

A great deal of theoretical work has been devoted to modeling the AGN feedback, 
including analytic and empirical models \citep[e..g.][]{Silk98, Fabian1999, Chen2020}, 
semi-analytic models \citep[e.g.][]{Croton2006, Bower2006}, and hydro-dynamical 
simulations \citep[e.g.][]{DiMatteo2005, Hopkins2007a, Schaye2015, Dubois2016, 
Pillepich2018, Dave2019}. Many of the mechanisms proposed are relevant 
not only to SMBH and AGN formation
\cite[][]{DiMatteo2005, Hopkins2011}, 
but can also change the structure of the 
host galaxy, thus predicting some correlations of SMBHs (and 
their feedback) with the structure of host galaxies. Broadly, 
the mechanisms can be divided into two categories: 
rapid and secular \citep{Kormendy2004}.
Major mergers of galaxies, which belong to the first category, 
can scramble galaxy disks and trigger starbursts, 
produce a dynamically hot stellar system, such as a bulge or 
an elliptical. Bar-driven processes, which belong to the second 
category, can also rearrange stars and gaseous clouds in a galaxy.
In particular, bars can drive gas into the center of a galaxy and 
cause the buildup of a central concentration resembling a dynamically 
hot compact object. Interactions with neighboring 
galaxies and minor mergers may also play a role by pumping 
energy into a galaxy disk, making it dynamically hotter. 
These mechanisms can thus help us to understand the observed BHSRs \citep[see discussion in][]{Croton2006, Saglia2016ApJ, Graham2023}.

The Sloan Digital Sky Survey (SDSS) Mapping Nearby Galaxies at Apache Point Observatory \citep[MaNGA,][]{Bundy2015} project provides spatially resolved 
spectra of about 10,000 galaxies and offers an unprecedented opportunity 
to study star formation quenching and its relation to galaxy properties. 
The data can be used to obtain spatially resolved information 
about quenching and dynamical properties within individual 
galaxies, thus allowing more comprehensive analyses on the connection 
between quenching and structural and dynamical properties of galaxies.  
For example, recent studies based on MaNGA showed that quenching is inside-out, 
namely starts from the central region and spreads outwards 
\citep[e.g.][]{LiC2015ApJ, WangE2018a, Ellison2018, Bluck2020B}. Some galaxies are then expected to be quenched in central regions 
but star-forming in the outer parts. Investigating the connection 
between quenching and structural/dynamical properties 
in these hybrid galaxies is particularly valuable. 
Furthermore, we can use the resolved spectroscopy to derive 
velocity dispersion in different parts of individual galaxies
and use it as a parameter to establish a link to the 
observed BHSRs. 

The paper is organized as follows. Section \ref{sec_data} briefly 
describes MaNGA, our sample selection and data reduction, in particular 
the method to define galaxies/objects of different quenching properties. 
In Section \ref{sec_dhot}, we analyze the data and show that fully 
quenched galaxies and quenched central cores both follow the 
quenched galaxy scaling relation. We also investigate the dynamical status 
using the distribution of galaxies in space spanned by  
velocity dispersion measured in the inner and outer regions of galaxies. 
Section \ref{sec_coe} presents discussions on an evolutionary 
track in dynamical status and quenching. We also discuss potential 
processes that can both change the dynamical status of galaxies and 
trigger AGN activities in them. 
In Section \ref{sec_QGBH}, we construct a toy model to connect 
quenching, the quenched galaxy scaling relation, BHSRs and AGN feedback. 
Finally, we summarize our results in 
Section \ref{sec_sum}. Throughout the paper, we adopt the following 
cosmological parameters: $H_0=73{\rm km/s/Mpc}$, $\Omega_{\rm M}=0.3$
and $\Omega_{\rm \Lambda}=0.7$.

\section{Data and analyses}\label{sec_data}

We describe galaxy samples used in our analyses in 
Section \ref{sec_sample}, and global and spatially resolved parameters 
of of individual galaxies in Section \ref{sec_pipe3d}. 
Section \ref{sec_gclass} shows how we classify galaxies based on 
their spatially resolved quenching properties.  
Section \ref{sec_fitting} describes a linear regression 
method used to analyze the observational data. 
Key symbols and terminologies used in this paper are listed 
in Table \ref{tab_term} for reference.

\subsection{MaNGA galaxies and sample selection}\label{sec_sample}

Our galaxy sample is taken from Mapping Nearby Galaxies at Apache Point Observatory \citep[MaNGA,][]{Bundy2015} of Solan Digital Sky Survey Data Release 17 
\citep[SDSS DR17,][]{Abdurrouf2022}. MaNGA galaxies are observed by the 
integral field units (IFUs) which have a wavelength coverage from 
3600 $\rm \AA$ to 10000 $\rm \AA$ and a spectral resolution of about 2000.  
This enables us to analyze spatially resolved quenching and dynamical properties 
of galaxies.
Over 10,000 galaxies have been observed during the period from April 2014 
to August 2020, and the main MaNGA sample is selected to have a flat 
stellar mass distribution \citep{Wake2017AJ}. 
The total data set consists of three sub-samples: Primary, Secondary and 
Color-Enhanced Samples, with Color-Enhanced Sample being a supplement 
of galaxies in the green valley. For Primary and Color-Enhanced Samples, 
the field of view (FOV) covers an aperture of $1.5R_{\rm e}$ or larger, 
where $R_{\rm e}$ is the effective (or half-light) radius of the galaxy, 
while the FOV of Secondary Sample covers an aperture of $2.5R_{\rm e}$
or larger. For the sake of comparison, our analyses are mainly based 
on galaxy properties within $1.5R_{\rm e}$.

 We adopt the data products of Pipe3D\footnote{\urlstyle{rm}\url{https://data.sdss.org/sas/dr17/manga/spectro/pipe3d/v3_1_1/3.1.1/}}
\citep{Sanchez2016RMxAA1, Sanchez2016RMxAA2, Sanchez2022ApJS}, 
which provides data for 10,220 galaxies from the analyses of 
10,245 data cubes with 25 failed cases (some more information about 
the Pipe3D is given in the following subsection).
MaNGA did some repeated observations 
for spectrophotometric calibrations. For such cases, we choose data
products with the highest signal-to-noise ratio (S/N) in the stellar
mass ($M_*$) measurement. We remove galaxies that do not have valid 
NSA redshifts or $R_{\rm e}$, which results in a total of 9,988 galaxies. 
We also apply two further restrictions: (i) excluding galaxies 
with QCFLAG equal to 1 or 2, i.e. galaxies with wrong redshift or low S/N, 
or with no value listed for QCFLAG from spectral fitting;
(ii) galaxies are required to have valid measurements for the central velocity dispersion ($\sigma_{\rm in}$), velocity dispersion in the outskirts ($\sigma_{\rm out}$) and the 
star formation rate (SFR). With these two restrictions, a total of 
9,591 galaxies remain. 

To have a galaxy sample that may be less affected by environmental processes,  
we focus on central galaxies in galaxy groups and clusters.  
To achieve this, we cross-match the Pipe3D catalog with the SDSS group 
catalog provided by \cite{Yang2007ApJ}. We obtain 8,888 galaxies, among 
which 6,231 are centrals.  
We only consider galaxies with $\log M_*/\Msun \geq 10$, 
because star formation quenching plays a less important role in 
these low-mass centrals. Our final central galaxy sample consists 
of 5,124 galaxies, with stellar mass covering the range 
$10^{10.0}-10^{12.2}\Msun$.  
  
Our analyses also use optically-selected AGNs, radio galaxies and barred galaxies. To obtain an optical AGN
subsample, we first cross-match our sample galaxies with the MPA-JHU 
catalog \citep{Brinchmann2004,Kauffmann2003mpajhu} 
to obtain the fluxes in four emission lines: ${\rm H\alpha}$, ${\rm H\beta}$, 
${\rm [NII\lambda6584]}$ and ${\rm [OIII\lambda5007]}$. We then use the 
the BPT diagnostic diagram to identify 578 galaxies that 
are classified as optical AGNs according to the criterion proposed by \cite{Kewley2001}. 
We cross-match our galaxies with the radio source catalog provided by \cite{Best2012} and obtain 156 radio galaxies.
To select barred galaxies, we cross-match our sample galaxies 
with the Galaxy Zoo DR2 (GZ2) catalog \citep{Willett2013MNRAS,Hart2016MNRAS}, 
and identify barred galaxies as the ones with 
\texttt{t03\_bar\_a06\_bar\_flag}=1, as recommended by GZ2. 
This yields a subsample of 421 barred galaxies.  
In the following, we also investigate the $r$-band S$\rm \Acute{e}$rsic index, $n$, taken from the NYU value-added galaxy catalog \citep[NYU-VAGC,][]{Blanton2005}.

\subsection{Measurements from Pipe3D}\label{sec_pipe3d}

\begin{table}
\caption{Terminologies and symbols}
\label{tab_term}
\centering
\begin{tabular}{c c}
\hline  
Terminology &  Criterion and description\\
\hline  
FQG       &  $\bar f_{\rm q}>95\%$, fully quenched galaxy\\
PQG       &  $5\%\leq\bar f_{\rm q}\leq95\%$, partically quenched galaxy\\
FSG         &  $\bar f_{\rm q}<5\%$, fully star-forming galaxy\\
QCC        &  $\bar f_{\rm q}=95\%$, $R_{\rm q}>4''$, quenched central core\\
SRDH  &  scaling relation of dynamical hotness\\
QGSR  &  quenched galaxy $M_*$-$\sigma_{\rm in}$ scaling relation\\
BHSR  & supermassive black hole scaling relation\\
\hline 
\end{tabular}
\begin{tabular}{c c}
Symbols & Description \\
\hline
$M_{*}$     & stellar mass of a galaxy in FOV\\
$R_{\rm e}$     & $r$-band half-light radius of a galaxy or QCC\\
$n$ & $r$-band S$\rm \Acute{e}$sic index of a galaxy\\
sSFR & $\rm H\alpha$-based specific star formation rate in FOV\\
$M_{\rm q}$     & stellar mass of a QCC\\
$R_{\rm q}$     & size of a QCC\\
$\bar f_{\rm q}$     &  quenched fraction in $1.5R_{\rm e}$ or in $R_{\rm q}$\\
$\sigma(R)$     & velocity dispersion profile\\
$\sigma_{\rm in}$     & median $\sigma$ of spaxels within $0.2R_{\rm e}$\\
$\sigma_{\rm out}$     & median $\sigma$ of spaxels in $1.4R_{\rm e}<R<1.5R_{\rm e}$\\
$\Sigma_{1}$ & surface mass density in central $1\,\rm kpc$\\
$\sigma_{\rm hot}$ &  $\log\sigma_{\rm hot}=0.3\log M_* -1.1$(SRDH, Eq. \ref{eq_fjr})\\
$\sigma_{\rm q}$ &  $\log\sigma_{\rm q}=0.30\log M_* -1.02$(QGSR, Eq. \ref{eq_qgsr})\\
$\alpha$ & $\alpha=0.30$, the slope of the QGSR  \\
$\beta$ & $\beta=-1.02$, the intercept of the QGSR \\
$\Delta_{\rm q}$ & $\Delta_{\rm q}=0.08$, the intrinsic scatter of the QGSR\\
\hline
\end{tabular}
\end{table}

Pipe3D provides a comprehensive analysis of both stellar continuum and gas 
emission lines. For our study, we use the newly released data products 
derived from \texttt{pyPipe3D} \citep{Sanchez2022ApJS}. Pipe3D analyzes 
the data cubes sampled linearly in wavelength and pre-processes them with 
Galactic extinction and a normalised spectral resolution. 
Due to a minimum S/N requirement for the analysis, contiguous spaxels 
are grouped together to make a spatial bin (tessella). 
Pipe3D first performs a simple stellar population (SSP) fitting 
to the stellar continuum. The SSP spectrum is shifted 
according to the mean stellar velocity,   
multiplied by the Cardelli extinction curve \citep{Cardelli1989ApJ}
to account for dust extinction, and convolved with a Gaussian function 
to account for stellar velocity dispersion ($\sigma$). 
The \texttt{MaStar\_sLOG} library \citep{Yan2019ApJ} sampled 
by 273 SSP spectra with 39 ages (from 1 Myr to 13.5 Gyr) and 7 
metallicities (from 0.006 $Z_{\odot}$ to 2.353 $Z_{\odot}$), and 
the Salpeter \citep{Salpeter1955ApJ} initial mass function (IMF), 
are adopted by Pipe3D. The modelled stellar spectrum is then 
subtracted from the observed one to obtain a `pure-GAS spectrum' 
that consists of emission lines, noise and residuals from the 
first (SSP-fitting) step. Strong emission lines in each 
spaxel are fitted with individual Gaussian functions. 
Pipe3D runs a series of Monte Carlo iterations with the input spectrum 
perturbed with its errors. The parameters quoted later are the mean values 
of the results from the iterations, and the corresponding errors are the 
standard deviations that account for both the noise and uncertainties 
in the fitting. 

Spatially resolved parameters are the direct products of Pipe3D. The stellar velocity 
dispersion ($\sigma$) for each spaxel is obtained simultaneously in the 
SSP fitting. The mass-to-light ratio and surface brightness can be used to 
estimate the surface mass density ($\Sigma_*$) of each spaxel. 
The $4000\rm \AA$ break (D4000) is defined as the flux ratio between 
the red and blue sides of $\lambda 4000\rm \AA$, as in  
\citet{Bruzual1983ApJ} and \cite{Gorgas1999AA}. This definition is less sensitive 
to smearing by velocity dispersion than the narrow version \citep{Sanchez2016RMxAA2,Bluck2020}. 
The amplitude of D4000 is 
believed to be a good indicator of stellar age 
\citep[see e.g.][]{Kauffmann2003mpajhu}, and we use it to define 
quenching properties for each spaxel (see Section \ref{sec_gclass}). 
Pipe3D provides two sets of mask extensions: the first is 
the GAIA\_MASK extension, which masks spaxels that are impacted 
by bright field stars; the second is the SELECT\_REG extension, 
which masks spaxels with low S/N spectra.


Pipe3D also provides global parameters for each galaxy. Some of
them are directly taken from the NSA catalog, such as redshift, 
effective radius ($R_{\rm e}$), position angle (PA) and ellipticity
\citep{Blanton2011AJ}. In Pipe3D, ellipticity 
is defined as $\sqrt{1-(b/a)^2}$, where $b$ and $a$ are the lengths of 
minor and major axes, respectively. The effective radius, $R_{\rm e}$, 
is defined as the Petrosian half-light radius in the $r$-band, 
and PA is defined as direction in which the major axis lies.
The ellipticity is an indicator of the inclination of a disk galaxy
and the shape of an elliptical galaxy. We use these parameters  
to estimate their profiles and other properties of individual galaxies.
The integrated quantities, such as stellar mass ($M_*$) and star formation rate (SFR), are also provided by Pipe3D 
by adding corresponding values of individual spaxels within the whole 
field of view (FOV) of the corresponding IFU, excluding masked spaxels. 
The dust-corrected ${\rm H}_{\rm \alpha}$ 
luminosity of a spaxel is used to estimate the star formation rate 
surface density ($\Sigma_{\rm SFR}$) using the relation given by \cite{Kennicutt1998}.
Pipe3D coadds the $\Sigma_{\rm SFR}$ within the FOV to derive the 
integrated star formation rate ($\rm SFR_{\rm H \alpha}$). Note that Pipe3D does 
not exclude ${\rm H}_{\alpha}$ emission that is not from young massive 
stars. The $\rm SFR_{\rm H \alpha}$ may thus represent an upper limit for 
some galaxies. Pipe3D also provided $\rm SFR_{\rm SSP}$ using three different 
timescales (10, 32 and 100 Myr) based on the SSP analysis. 
Tests show that these $\rm SFR$s are strongly correlated, although  
there are some systematic offsets. In this paper, we adopt the $\rm SFR$ 
estimated from the ${\rm H}_{\alpha}$ emission line. 
The specific star formation rate (sSFR) is defined as ${\rm SFR}/M_*$. 
All the parameters are listed in Table \ref{tab_term} along with their 
definitions.
 
 For values adopted directly from Pipe3D, we use the corresponding 
measurement errors provided by Pipe3D to represent the statistical 
uncertainties in our analyses. Pipe3D does not provide measurement errors 
for $R_{\rm e}$, and we follow \cite{Sahu2020ApJ} to assume a 30\% uncertainty 
for this quantity. This uncertainty for $R_{\rm e}$ is used in modeling  
the $M_*$-$R_{\rm e}$ relation in Section \ref{sec_QCCs}, and our tests show that 
the the result is not sensitive to the choice of the uncertainty.

 To derive parameters from the spatially resolved 
products of Pipe3D, it is necessary to check the quality of 
the data for this purpose. Among all spaxels within $1.5R_{\rm e}$ of 
our selected galaxies, about 0.14\% and 3.99\% are in 
GAIA\_MASK and SELECT\_REG masks, respectively.  
Our tests show that Pipe3D products for most of the 
unmasked spaxels have relatively good qualities. 
Among all the 5,124 galaxies, 4,978 of them have 
all their spaxels in $1.5 R_{\rm e}$ with $S/N>5$ in D4000;   
4,387 galaxies have all spaxels with $S/N>3$ in $\Sigma_*$;
and 3,509 galaxies have all spaxels with $S/N>3$ in $\sigma$. 
This indicates that $\sigma$ is more sensitive to 
spectral qualities than the other two quantities. 
Our further test shows that, in 80\% of the sample galaxies, more 
than 85\% of their spaxels have $S/N>3$ in $\sigma$. 
In our analyses, we exclude spaxels with $S/N\leq 3$ in $\sigma$. 

  We derive the profile of a parameter using selected spaxels within the central 
$1.5 R_{\rm e}$, and the observed PA and ellipticity are used to make the 
inclination correction \citep[see e.g.][]{Bluck2020B}.
For example, for each galaxy, we use the median value of $\sigma$
in an inclination-corrected annulus to calculate the $\sigma$ profile.
The related global parameter of the galaxy is obtained from the profile. 
For instance, the central velocity dispersion ($\sigma_{\rm in}$) is defined 
as the median value of $\sigma$ within the central $0.2 R_{\rm e}$ 
and the velocity dispersion in the outskirts ($\sigma_{\rm out}$)
is the median $\sigma$ within 1.4-1.5$R_{\rm e}$. 
The central stellar mass surface density ($\Sigma_1$) is calculated by 
first coadding the $\Sigma_*$ of spaxels within the central 1 kpc 
(also inclination corrected) and then dividing it by the summation of 
the physical areas of the spaxels. The profiles of the quenched fraction 
shown in the next section are also calculated using inclination-corrected 
annuli. We list these parameters and their descriptions in 
Table \ref{tab_term} for reference.

For a quantity, $S$, derived from the measurements of a set 
of individual spaxels, we use the following procedure to obtain 
its uncertainty. For each spaxel in the set, the measurement of 
the relevant spaxel quantity, $s$, and the corresponding uncertainty, $e_s$, 
both given by Pipe3D, are used to generate a new $s'$ from 
$N(s, e^2_s)$ using a Monte Carlo method. The set of the 
Monte-Carlo generated $s'$ values are then used to derive an
$S'$ in the same way as $S$ is obtained. We repeat this process 500 times to obtain 
the distribution of $S'$ and use the standard deviation of the 
distribution to represent the uncertainty of the derived quantity. 
For example, $\sigma_{\rm in}$ and $\sigma_{\rm out}$ are defined as the 
median values of the $\sigma$ in the corresponding sets of spaxels, 
so that their errors are obtained from the variances of the medians of the 
500 Monte Carlo samples. We note again that this procedure is valid only under 
the assumption that the uncertainties in different spaxels are independent.

\subsection{Classification of galaxies}\label{sec_gclass}

\begin{figure*}[htb]
    \centering
    \includegraphics[scale=0.56]{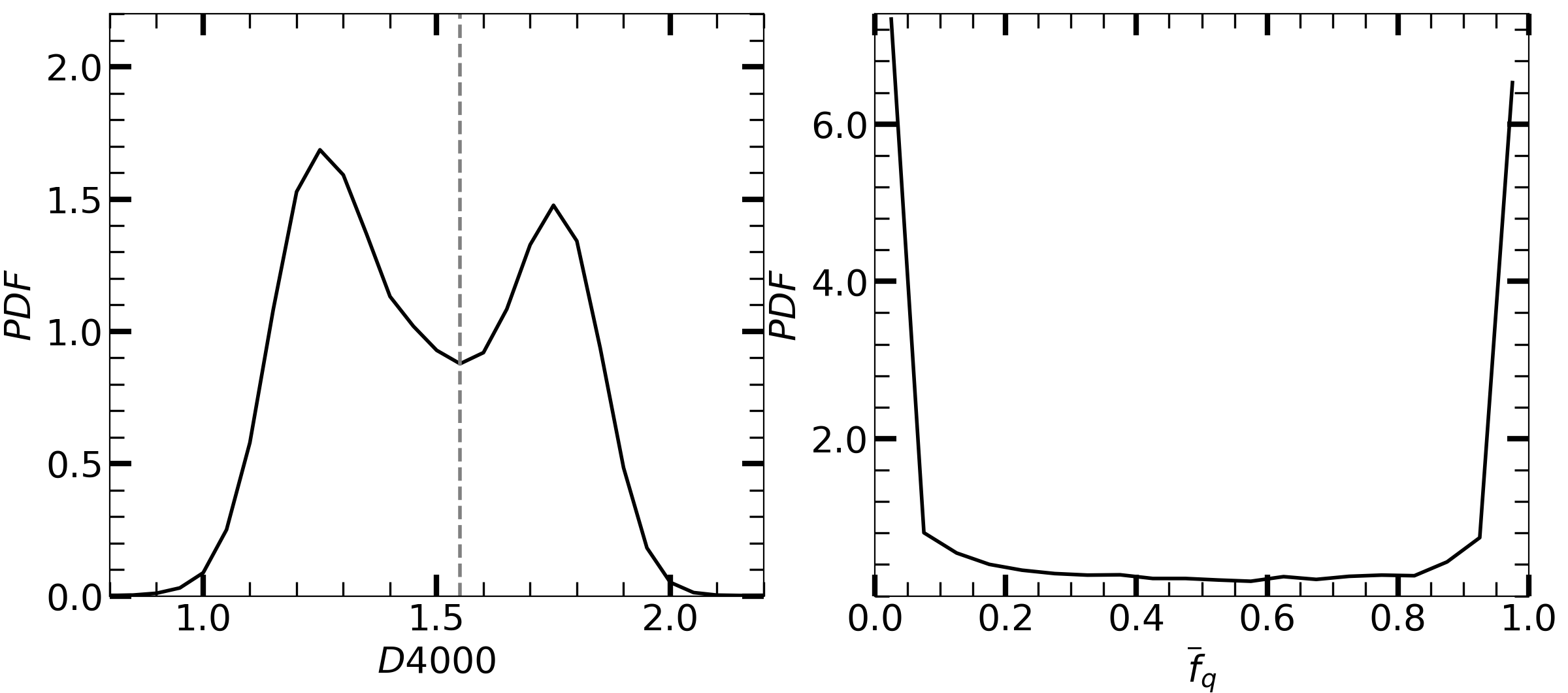}
    \caption{Probability distribution functions of spatially resolved D4000 (left panel) and mean quenched fraction ($\bar f_{\rm q}$, right panel) for our galaxy sample. In the left panel, we show the result for D4000 of spaxels within 1.5$R_{\rm e}$ and the grey dashed line indicating $\rm D4000_{\rm th}=1.55$. In the right panel, $\bar{f}_{\rm q}$ is calculated with $\rm D4000_{\rm th}=1.55$.}
    \label{fig_pdf_d4000_meanfq}
\end{figure*}

As shown in the left panel of Figure \ref{fig_pdf_d4000_meanfq}, 
the D4000 distribution 
of spaxels is bimodal,  with a valley around 1.55.
Since the D4000 of a spaxel is strongly correlated with its  
SFR surface density \citep[e.g.][]{Brinchmann2004}, we use 
$\rm D4000_{\rm th}=1.55$ to separate spaxels into two categories, 
quenched and star-forming. We define the quenched fraction, 
$f_{\rm q}$, as the fraction of spaxels with $\rm D4000>D4000_{\rm th}$.
For each galaxy, we calculate the mean quenched fraction, 
$\bar f_{\rm q}$, within $1.5R_{\rm e}$, as well the quenched profile, 
$f_{\rm q}(R)$, which is the quenched fraction as a function of the
radius to the galaxy center. The ${\bar f}_q$ distribution of galaxies, 
plotted in the right panel of Figure \ref{fig_pdf_d4000_meanfq}, 
peaks towards the two extremes, suggesting that quenching usually occurs over the 
entire galaxy \citep[see also,][]{WangE2018a, Bluck2020}. Based on this distribution, we classify galaxies 
into several classes in a way similar to that proposed in \cite{WangE2018a}.
Fully quenched galaxies (FQGs) are defined as the ones with 
$\bar f_{\rm q}>\bar f_{\rm q, th}=0.95$. In this class, 
the majority of the spaxels within $1.5R_{\rm e}$ are quenched spaxels.
Fully star forming galaxies (FSGs) are defined as the ones 
that have $\bar f_{\rm q}<1-\bar f_{\rm q, th}=0.05$, 
i.e. the majority of their spaxels are classified as star-forming spaxels. 
The rest of the galaxies are classified as partially quenched galaxies 
(PQGs). This classification results in 1,672 FQGs, 1,879 FSGs, and 1,573 
PQGs. The acronyms are listed in Table \ref{tab_term} for reference.

\begin{figure}
    \centering
    \includegraphics[scale=0.45]{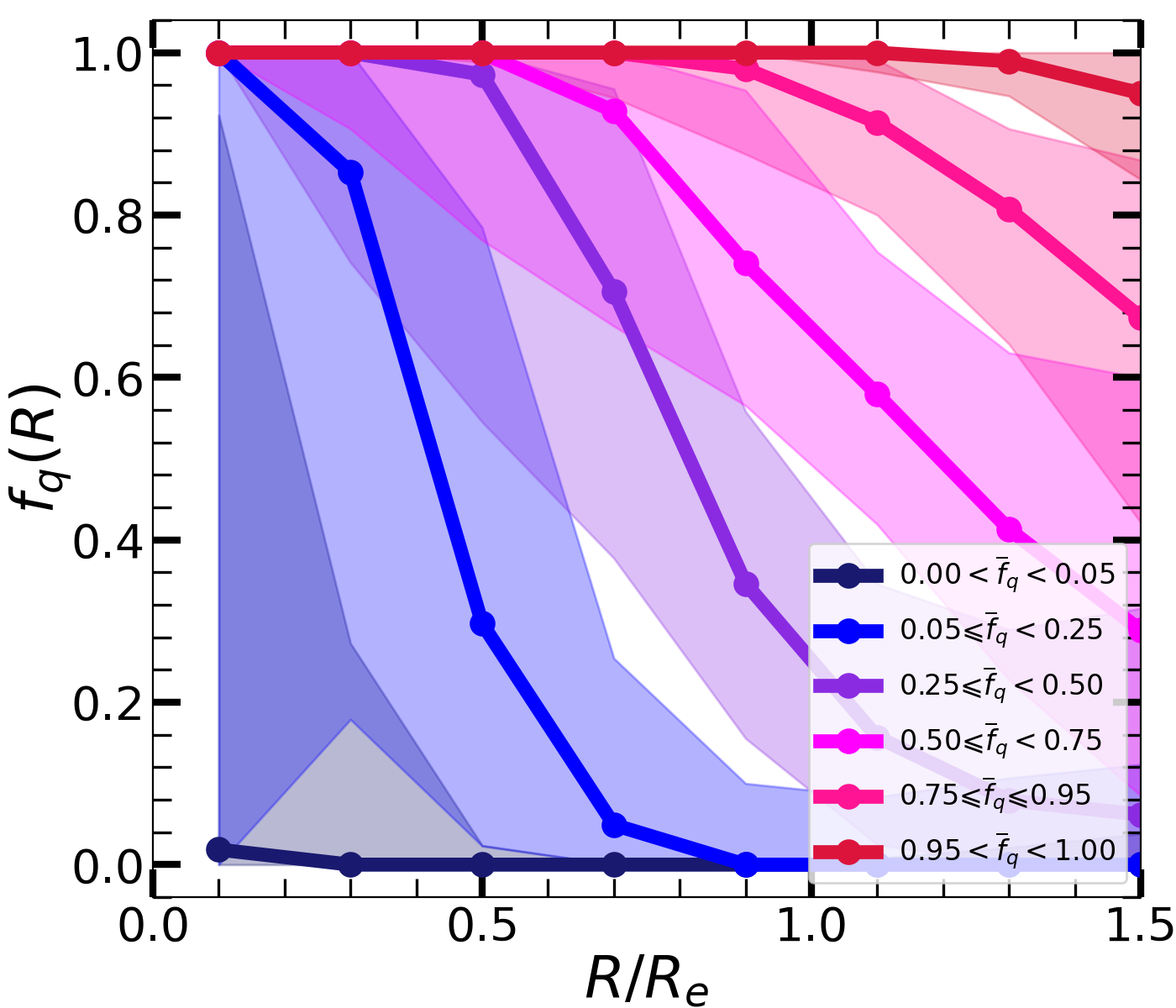}
    \caption{The median quenched profiles ($f_{\rm q}(R/R_{\rm e})$) for galaxies with different $\bar{f}_{\rm q}$ bins as indicated in the figure. The shadows show the 16th and 84th percentiles of the distribution of $f_{\rm q}$. Note that galaxies with $\bar f_{\rm q}<0.05$ are FSGs (dark blue), galaxies with $\bar f_{\rm q}>0.95$ are FQGs (red), the rest galaxies are PQGs. We do not show galaxies with $\bar f_{\rm q}=0$ or $\bar f_{\rm q}=1$.}
    \label{fig_prof_fq}
\end{figure}

The median quenched profiles for galaxies in different bins of 
$\bar f_{\rm q}$ are shown in Figure \ref{fig_prof_fq}. We can see that 
$\bar f_{\rm q}$ is about one at small radii for all cases, 
and drops at larger radii. The drop is quite rapid for galaxies 
with low $\bar f_{\rm q}$. This suggests that a PQG usually has a 
fully quenched core in its center. We can thus define a 
minimum radius, within which the quenched fraction is equal to $\bar f_{\rm q, th}$. We refer to this radius as the quenched radius, denoting it as $R_{\rm q}$. The mass $M_{\rm q}$ of the 
quenched central core (QCC) is defined as the sum of the stellar mass within 
$R_{\rm q}$. We assume that the fractional errors in $M_{\rm q}$ and 
$R_{\rm q}$ are the same as $M_*$ and $R_{\rm e}$, respectively. 
In the calculation, a spaxel without $\Sigma_*$ 
measurement is given the median value of the spaxels that have 
the same radius as the spaxel. We only consider QCCs with 
$R_{\rm q}\geq 4''$ to ensure that the cores can be well resolved 
and our measurements are not significantly contaminated by 
structures larger than $R_{\rm q}$. We also try to identify QCCs 
in FSGs with non-zero $\bar f_{\rm q}$, but they are too small
to meet the criterion of $R_{\rm q}\geq4''$. Thus, QCCs are identified only 
in PQGs. For each QCC, we define its $\sigma_{\rm in}$ as the median 
$\sigma$ of the spaxels within $0.2R_{\rm e}$ 
and $\sigma_{\rm out}$ as the median $\sigma$ in 1.4-1.5$R_{\rm e}$, 
where $R_{\rm e}$ for QCCs are defined in Section \ref{sec_QCCs}. 
Sixty one QCCs do not have valid $\sigma_{\rm in}$ or $\sigma_{\rm out}$ measurements 
and are excluded. This leaves a total of 599 QCCs for our analyses.

We have performed a series of tests with various $\rm D4000_{\rm th}$
and $\bar f_{\rm q, th}$ to check whether our results are sensitive to 
the choices of these thresholds. These tests are based on 
nine combinations of three choices of $\rm D4000_{\rm th}$ 
(1.5, 1.55 and 1.6) and three choices of $\bar f_{\rm q, th}$ 
(0.9, 0.95 and 0.98). Most of the results shown in this paper 
are affected only slightly, and our conclusions remain unchanged. 
As shown in Figure \ref{fig_pdf_d4000_meanfq}, D4000 has a bimodal 
distribution, and a small variation of $\rm D4000_{\rm th}$ around 
the valley ($\sim1.55$) does not lead to a significant change in 
the sample classification. The distribution of ${\bar f}_{\rm q}$, 
shown in the right panel of Figure \ref{fig_pdf_d4000_meanfq}
using $\rm D4000_{\rm th}=1.55$, is also strongly bimodal, and so 
a small change in $\bar f_{\rm q, th}$ does not lead to any significant  
changes in our conclusions.

\begin{figure*}[htb]
    \centering
    \includegraphics[scale=0.39]{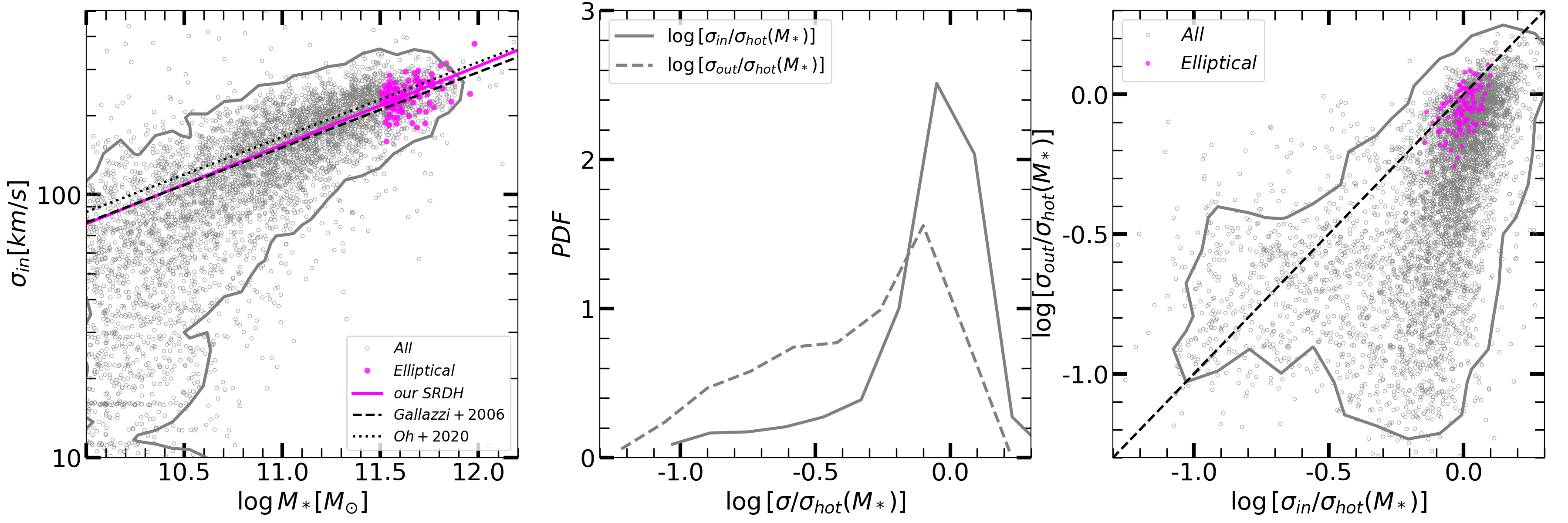}
    \caption{Left panel: $\sigma_{\rm in}$ versus $M_*$. The pink solid line represents the SRDH (Equation \ref{eq_fjr}). Its slope is fixed to 0.3 and its amplitude is determined by massive ($M_*>10^{11.5} M_{\odot}$) elliptical galaxies (pink solid circles). Grey circles show all galaxies used in this paper and grey line corresponds to the 95\%-th contour level. The black dashed \citep{Gallazzi2006}, dotted \citep{Oh2020} lines are the Faber-Jackson relation taken from literature for comparison. Middle panel: Probability distribution functions of $\log(\sigma_{\rm in}/\sigma_{\rm hot})$ (grey solid line) and $\log(\sigma_{\rm out}/\sigma_{\rm hot})$ (grey dashed line) for all galaxies. Right panel: the Two-$\sigma$ diagram, $\sigma_{\rm in}/\sigma_{\rm hot}$ versus $\sigma_{\rm out}/\sigma_{\rm hot}$. Symbols are the same as that in the left panel except that the black dashed line shows the 1:1 relation.}
    \label{fig_FJR_ms_sig_2sig}
\end{figure*}

\begin{figure*}[htb]
    \centering
    \includegraphics[scale=0.38]{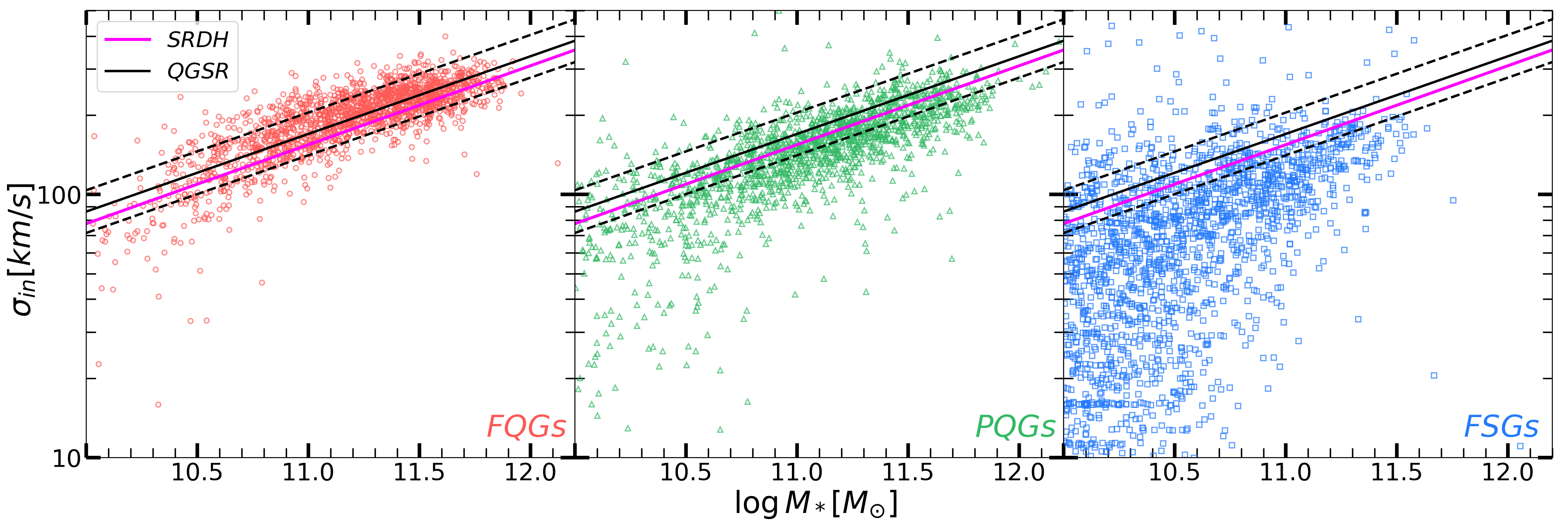}
    \caption{Central velocity dispersion ($\sigma_{\rm in}$) versus stellar mass ($M_*$) for three different classes of galaxies, FQGs (left: red circles), PQGs (middle: green triangles) and FSGs (right: blue squares). The pink solid line is the SRDH (Equation \ref{eq_fjr}). The black solid line shows the best fitting  for FQGs (Equation \ref{eq_qgsr}), i.e. quenched galaxy scaling relation (QGSR).  The black dashed lines represent the 1-$\Delta$ intrinsic scatter of this relation. 
    }
    \label{fig_sigc-ms}
\end{figure*}

\subsection{Correlation analyses}\label{sec_fitting}


Our investigation also uses correlations between galaxy properties 
described above, and here we describe the related analyses. 
Suppose that we have two series of data points, $X=\{x_1,...,x_n\}$
and $Y=\{y_1,...,y_n\}$, with the elements, $x_i$ and $y_i$,  
representing two properties of galaxy $i$ in the sample of $n$ galaxies.   
The uncertainties of the measurements are represented 
by the standard deviations, $e_{xx,i}$ and $e_{yy,i}$.
In many cases, we assume a linear relation between 
$Y$ and $X$, and we write 
\begin{equation}
    Y=a_1 X +a_2 + N(0,\Delta^2),\\\label{eq_linear}
\end{equation}
where $a_1$ is the slope, $a_2$ is the intercept, 
and we model the variance in the relation by 
$N(0,\Delta^2)$, a normal distribution with mean equal to 0 and 
variance equal to $\Delta^2$. Thus, $\Delta$ describes 
the intrinsic scatter of the relation in $Y$. 
To constrain the three free parameters, $a_1$, $a_2$ and $\Delta$, we 
construct a posterior distribution as,
\begin{equation}
    Posterior= Likelihood \times Prior. \label{eq_posterior}
\end{equation}
Taking into account of measurement uncertainties and 
intrinsic scatter,  
we write the likelihood term as,
\begin{equation}
\begin{split}
    Likelihood=\prod \limits_{i=1}^n \frac{1}{\sqrt{2\pi (e_{yy,i}^2+a_1^2 e_{xx,i}^2+\Delta^2)}}\times 
    \\
    \exp\{-\frac{[y_i-(a_1 x_i+a_2)]^2}{2(e_{yy,i}^2+a_1^2 e_{xx,i}^2+\Delta^2)}\}.\label{eq_likelihood}
\end{split}
\end{equation}
Given that we have little prior information about the free parameters, 
we take uninformative $prior$s, i.e. a flat distribution in a given 
range. We use the python code \texttt{emcee} 
\citep[]{Goodman2010CAMCS, Foreman013PASP}, which is the implementation 
of the ensemble sampler for Markov Chain Monte Carlo (MCMC), to sample the 
distribution of the model parameters.

\section{Dynamical hotness and star formation quenching}
\label{sec_dhot}


\subsection{Defining dynamical hotness}

As discussed above, the quenching status of a galaxy is correlated with its 
dynamical state. Here we define the dynamical hotness of a galaxy to 
characterize the randomness of stellar motion. 
Historically, galaxies obeying the Faber-Jackson relation (FJR) are called 
hot dynamical systems \citep[see e.g.][]{Bender1992}, and we adopt 
a similar relation to define dynamical hotness. 
Figure \ref{fig_FJR_ms_sig_2sig} shows $\sigma_{\rm in}$ versus $M_*$ 
for all our sample galaxies. For comparison, we also show the FJRs obtained by \cite{Gallazzi2006} and \cite{Oh2020}. The slopes of the two FJRs are 
very similar, 0.286 and 0.285, respectively, and the zero points differ 
by less than 0.05 dex in $\sigma_{\rm in}$. 
One can see, both FJRs describe well the trend followed by the 
hottest population with the highest $\sigma_{\rm in}$ at given $M_*$
in our MaNGA sample. However, the two FJRs were obtained using 
$M_*$ and $\sigma_{\rm in}$ measured in different ways from ours. 
It is thus necessary to re-calibrate the relation for our purpose.

Pipe3D provides morphology classification using a machine learning method 
based on several galaxy features, including S$\Acute{e}$rsic index, 
stellar mass, line-of-sight $V-\sigma$ ratio, $R_{\rm e}$, ellipticity, 
concentration and $ugriz$ colors \citep{Sanchez2022ApJS}. 
We select galaxies identified as ellipticals to represent dynamically 
hot galaxies. However, our visual inspection shows that some of the 
elliptical galaxies listed as Pipe3D are actually face-on spirals and 
have very small $\sigma_{\rm in}$. This mis-classification is severer 
for lower-mass galaxies, and the classification is more reliable 
at the massive end. We thus decide to use massive galaxies for our calibration. 
There are 158 galaxies with $\log{M_*/\Msun}>11.5$ classified as
ellipticals. We remove galaxies with close companions and/or significant 
asymmetrical structure. The remaining 103 ellipticals are shown 
in Figure \ref{fig_FJR_ms_sig_2sig} as the pink points. We then use 
a linear relation with a slope of 0.3 to fit these elliptical galaxies.
We fix the slope because previous studies usually gave similar slopes
\citep[see e.g.][]{Gallazzi2006, Oh2020}. The scaling relation 
obtained is 
\begin{equation}
    \log\sigma_{\rm hot}=0.3\log M_*-1.1
    \,\label{eq_fjr}
\end{equation}
(see also Table \ref{tab_term}). 
This relation is shown in Figure \ref{fig_FJR_ms_sig_2sig} as the pink line. For clarity, we will refer to Equation \ref{eq_fjr} as the 
scaling relation of dynamical hotness (or SRDH).
As one can see, our SRDH is very close to the FJRs obtained previously. 
We have also performed the same fitting to the total 158 elliptical 
galaxies and obtained a very similar result. As one can also see, 
the SRDH basically defines an upper boundary of the velocity 
dispersion at a given stellar mass, as expected from a dynamically hot 
system in which the structure is supported predominantly by the 
random motion of stars. 

 In what follows, we use the deviation of a galaxy from the SRDH to define 
the dynamical hotness of a galaxy. Specifically, we use the ratio of the measured 
velocity dispersion of a galaxy to the value of $\sigma_{\rm hot}$ obtained  
from its stellar mass to characterize its hotness. 
The middle panel of Figure \ref{fig_FJR_ms_sig_2sig} shows the probability distributions of $\sigma_{\rm in}/\sigma_{\rm hot}$ and 
$\sigma_{\rm out}/\sigma_{\rm hot}$. We can see that 
$\log(\sigma_{\rm in}/\sigma_{\rm hot})$ peaks around zero, with an 
extended tail down to $-1$. 
Compared to $\sigma_{\rm in}/\sigma_{\rm hot}$, $\sigma_{\rm out}/\sigma_{\rm hot}$ shows a much more extended 
distribution and a much lower peak, indicating that 
the dynamical hotness may depend on the distance to the center 
in an individual galaxy. Clearly, the dynamical hotness in outer part of a galaxy
provides additional information about the dynamical status of the galaxy. 
In the right panel of Figure \ref{fig_FJR_ms_sig_2sig} we show 
$\sigma_{\rm in}/\sigma_{\rm hot}$ versus $\sigma_{\rm out}/\sigma_{\rm hot}$ 
(hereafter, the Two-$\sigma$ diagram). As one can see, galaxy distribution 
in this diagram appears to be concentrated in two branches: 
a nearly horizontal branch in which $\sigma_{\rm out}/\sigma_{\rm hot}$ is low, and
a nearly vertical branch in which $\sigma_{\rm in}/\sigma_{\rm hot}$ is 
close to one. Elliptical galaxies used to calibrate the SRDH are  
located in a small region around $\log(\sigma_{\rm in}/\sigma_{\rm hot})\sim 0$ 
and $\log(\sigma_{\rm out}/\sigma_{\rm hot})\sim 0$, so that they  
are dynamically hot over their entire bodies. As we will see  
in Section \ref{sec_2sig}, galaxies of different quenching status show distinctive distributions 
in the Two-$\sigma$ diagram, making the diagram a powerful tool 
to understand quenching processes (see also Section \ref{sec_coe}).

\subsection{Dynamical hotness and quenching status}\label{sec_dhqs}

\begin{figure*}
    \centering
    \includegraphics[scale=0.29]{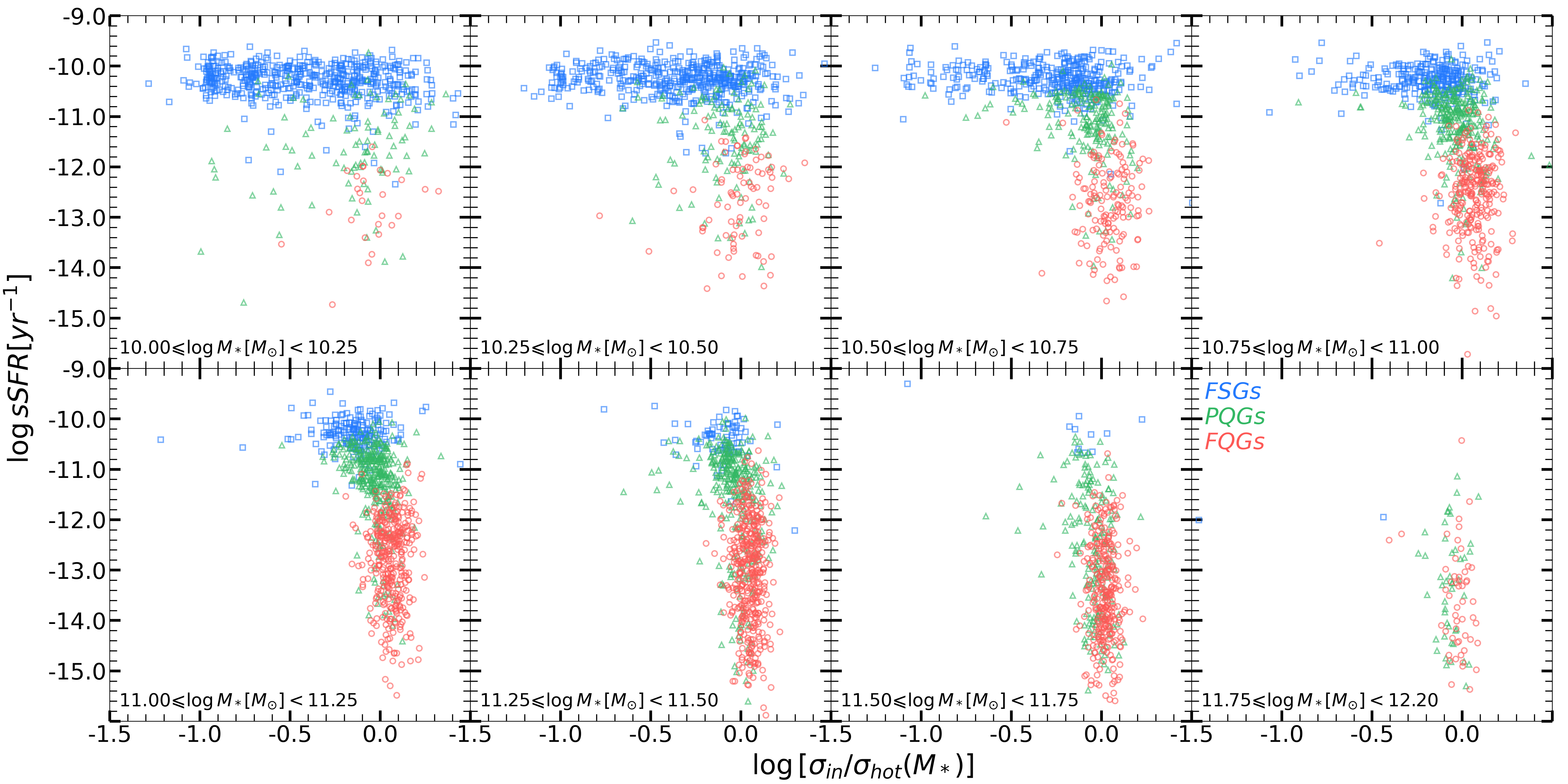}
    \caption{The sSFR as a function of $\log(\sigma_{\rm in}/\sigma_{\rm hot})$ for FQGs (red circles), PQGs (green triangles) and FSGs (blue squares). Galaxies of different stellar masses are shown in different panels.}\label{fig_ssfr-sigc}
\end{figure*}

Figure \ref{fig_sigc-ms} shows the central velocity dispersion versus the 
stellar mass for fully quenched galaxies (FQGs, left panel), partially 
quenched galaxies (PQGs, middle panel), and fully star-forming galaxies 
(FSGs, right panel). As one can see, FQGs exhibit a strong and tight 
correlation between $\sigma_{\rm in}$ and $M_*$. Using the same 
method described in Section \ref{sec_fitting}, we fit the relation
to a power-law function (Equation \ref{eq_linear}) 
and obtain
\begin{equation}
    \log\sigma_{\rm q}=\alpha \log M_*+(\beta \pm \Delta_{\rm q})\,,
    \label{eq_qgsr}
\end{equation}
with $\alpha=0.30$, $\beta=-1.02$, and $\Delta_{\rm q}=0.08$ (Table \ref{tab_term}).
This scaling relation is referred to as the quenched galaxy 
scaling relation (QGSR). The QGSR is similar to the relation for red 
galaxies obtained by \citet{Fang2013ApJ}. It is also very similar to 
the SRDH defined above to represent dynamically hot galaxies 
(Figure \ref{fig_sigc-ms}). This clearly shows that 
the FQGs are all dynamically hot in their central regions
(see also Section \ref{sec_2sig}). 

The $M_{*}-\sigma_{\rm in}$ relation for FSGs is very different 
from that for FQGs: at given $M_*$ the $\sigma_{\rm in}$ distribution 
is much broader. The majority of FSGs have lower $\sigma_{\rm in}$ than FQGs, 
in particular at small $M_*$. This is expected, as a large fraction of 
FSGs are spiral galaxies that have small bulges and low 
velocity dispersion in inner regions. A small fraction of FSGs 
are close to the SRDH, and some even have $\sigma_{\rm in}$ 
above the SRDH. As we will discuss in Section \ref{sec_coe}, 
the existence of these centrally hot star-forming galaxies 
has important implications for galaxy quenching and the related processes. 
The PQGs follow a $M_{*}-\sigma_{\rm in}$ relation that lies in between 
the FQG and FSG populations. For a given $M_*$, their $\sigma_{\rm in}$ 
is, on average, only slightly lower than the SRDH.  
A significant fraction of PQGs lie above the SRDH, indicating that they are dynamically hot in their 
inner regions.  The different $\sigma_{\rm in}$ distributions among the three populations indicate that quenching processes may be related to the 
velocity dispersion of the central stellar components, as suggested  
in some previous studies \citep[e.g.][]{Fang2013ApJ, Bluck2020, Brownson2022}. 

To see the relation between star formation and dynamical hotness more clearly, 
we show the specific star formation rate as a function of 
$\sigma_{\rm in}/\sigma_{\rm hot}$ for six narrow $M_*$ bins 
in Figure \ref{fig_ssfr-sigc}. As one can see, most of the FQGs have 
low star formation activities, with $\log \rm sSFR<-12$, and are 
dynamically hot in the inner region, 
with $\sigma_{\rm in}/\sigma_{\rm hot}\sim 1$. 
Note that the sSFR distribution of FQGs is very broad, ranging 
from $\log \rm sSFR=-12$ to $-15$, indicating that some of these 
galaxies may still have some low-level star formation activity
that is not correlated with $\sigma_{\rm in}$. 

In contrast, FSGs have a very narrow distribution in sSFR, 
with $-9.5>\log\rm sSFR >-10.6$, almost independent of stellar mass,  
indicating that they follow well the star-forming main sequence. 
The $\sigma_{\rm in}$-distribution of these galaxies is very broad, 
ranging from less than $0.1 \sigma_{\rm hot}$ to 
$\sim \sigma_{\rm hot}$, and the width of the distribution decreases with 
increasing stellar mass. Thus, FSGs of higher $M_*$, 
on average, follow the SRDH more closely. This is consistent with 
the fact that central components of spiral galaxies 
become more concentrated and bulge-dominated as $M_*$ increases 
\citep[see e.g.][]{Barro2017ApJ}. There is no significant correlation 
between $\sigma_{\rm in}$ and sSFR for FSGs. Overall, FSGs appear to 
form a horizontal sequence (in the $\sigma_{\rm in}/\sigma_{\rm hot}$-sSFR 
plane) that is almost perpendicular to the almost vertical sequence 
defined by the FQG population. 

\begin{figure}
    \centering
    \includegraphics[scale=0.48]{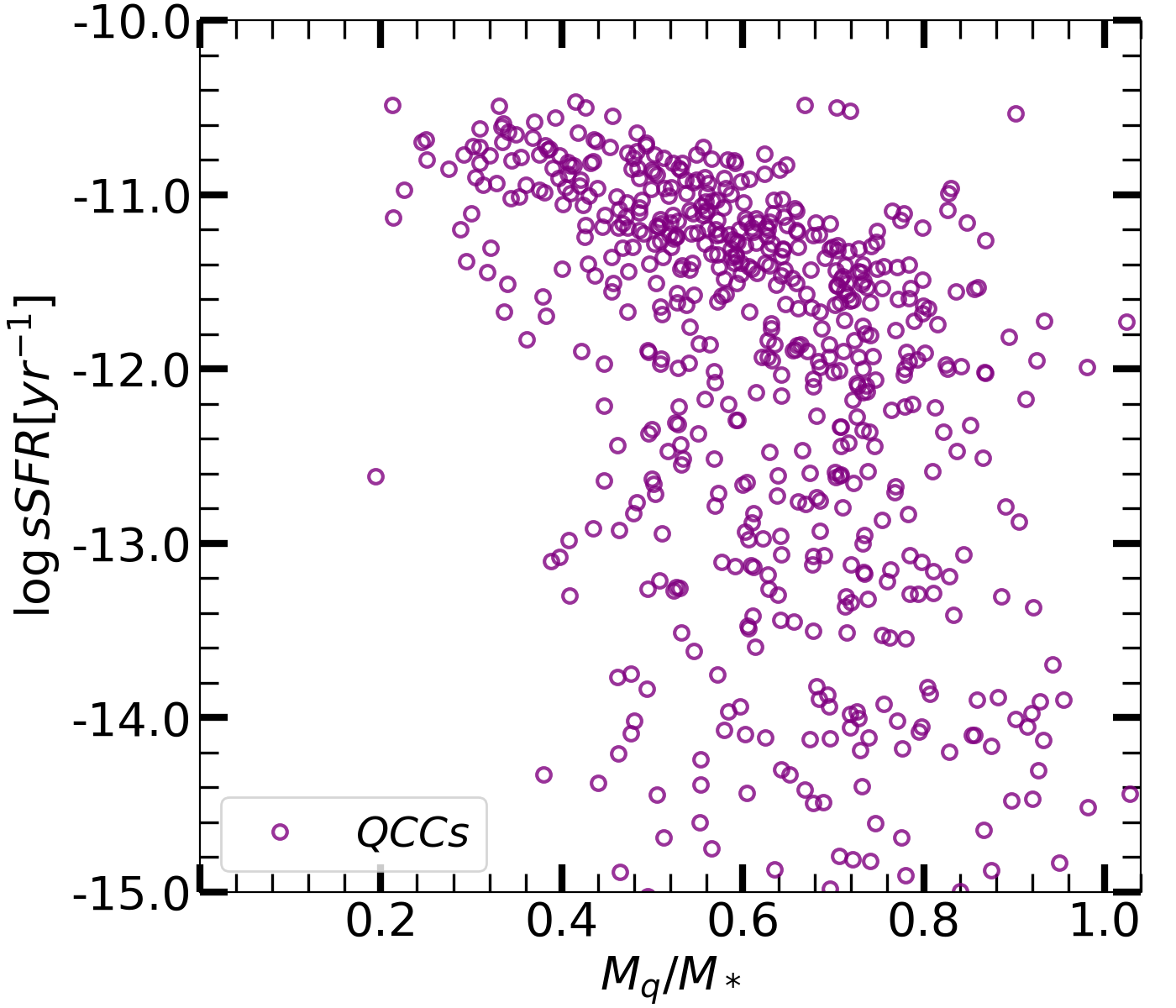}
    \caption{The sSFR versus $M_{\rm q}/M_*$ for QCCs. The sSFR is measured within the whole galaxy rather than within the QCC.}\label{fig_ssfr_mqms}
\end{figure}

\begin{figure}
    \centering
    \includegraphics[scale=0.45]{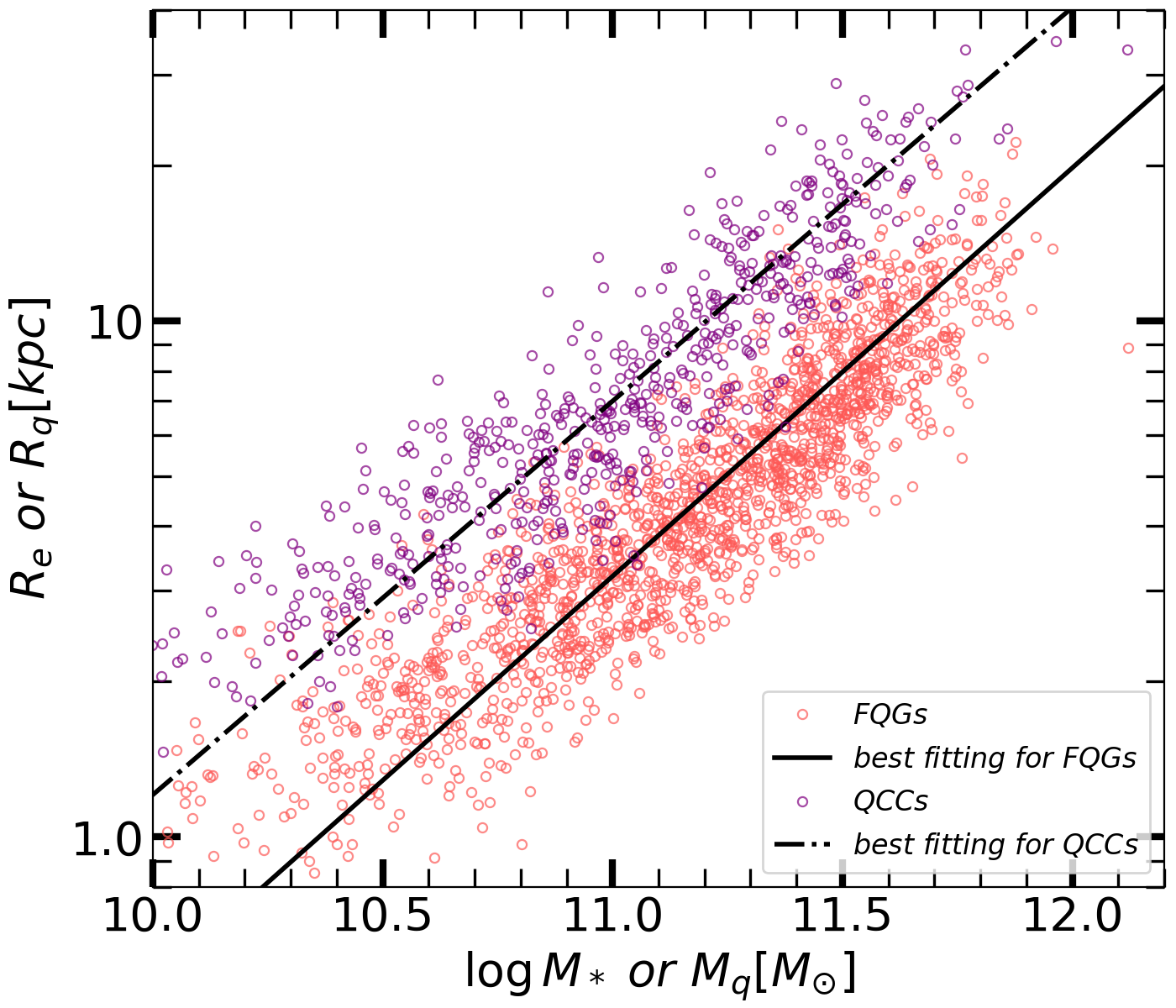}
    \caption{Red circles: $R_{\rm e}$ versus $M_*$ for FQGs. Purple circles: $R_{\rm q}$ versus $M_{\rm q}$ for QCCs. The solid and dash-dotted lines show the best-fitting results for FQGs and QCCs respectively. 
}\label{fig_mass_size}
\end{figure}

\begin{figure*}
    \centering
    \includegraphics[scale=0.39]{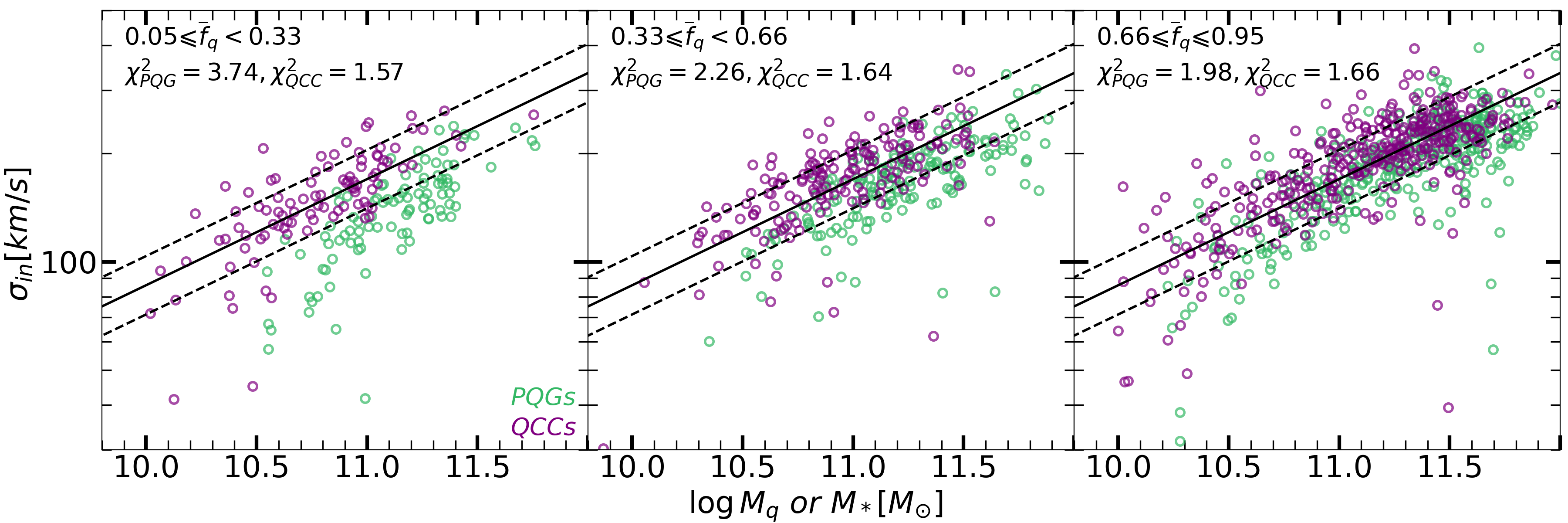}
    \caption{The $\sigma_{\rm in}$ versus $M_{\rm q}$ for QCCs (purple) and the $\sigma_{\rm in}$ versus $M_*$ for PQGs in which QCCs are identified. The black solid lines show the QGSR (Equation \ref{eq_qgsr}) and dashed lines show the intrinsic scatter. 
     We show galaxies in different $\bar f_{\rm q}$ bins in different panels. 
     The parameters $\chi^2$ presented in each panel are calculated using Equation \ref{eq_chi_QCC_PQG} and show the distances of PQGs and QCCs to the QGSR. Please see the text for details. 
     }\label{fig_ms_sigc_QCCs}
\end{figure*}

There is a clear gap between the two 
sequences defined by FQGs and FSGs, which is bridged by PQGs, 
as shown by the green points in  in Figure \ref{fig_ssfr-sigc}. The sSFR 
distribution of the PQG population is actually very broad, touching the 
sequence defined by FSGs and having a long tail extended well into 
the sequence defined by FQGs. PQGs with $\log\rm sSFR<-12$ have almost the same $\sigma_{\rm in}$ distribution as FQGs and their sSFR is also quite 
independent of $\sigma_{\rm in}$. These PQGs usually have large 
$\bar f_{\rm q}$, and overlap with the FQG population in many properties, 
as we will show later. These results suggest that PQGs with low sSFR 
are either on their way to become FQGs in the near future or are 
FQGs that have acquired some star-forming gas in the recent past.  
This sub-population consists of 375 galaxies, less than a quarter of the 
total PQG population. For the rest 3/4 of the PQG population
with $\log\rm sSFR>-12$, there is a noticeable negative trend 
of sSFR with $\sigma_{\rm in}/\sigma_{\rm hot}$ for $\log(M_*/\Msun)>10.5$. 
As shown in Section \ref{sec_QCCs}, many of these galaxies possess a 
quenched central core (QCC; see also Figure \ref{fig_prof_fq}), and the mass of the 
QCC increases with $\sigma_{\rm in}$.

 The three populations together form an L-shaped pattern in the 
 sSFR-$\sigma_{\rm in}$ plane. This L-shape is well-defined
in all stellar mass bins, except a tiny fraction of low mass galaxies 
residing at the lower-left corner (we will discuss these galaxies 
in Section \ref{sec_spe}). This is consistent with the 
results of \cite{Fang2013ApJ} and \cite{Barro2017ApJ} 
based on stellar mass surface densities. The L-shape suggests a 
transition in the evolution of star formation and its relation 
to dynamical hotness. It shows that a galaxy is 
fully quenched only when it has evolved to a dynamical state  
that is close to the SRDH. It is widely believed that $\sigma_{\rm in}$ 
is strongly correlated with the mass of the supermassive black hole (SMBH) 
residing in galaxy center \citep{Kormendy2013ARA&A}. 
Our results may thus suggest that a galaxy can be fully quenched 
only after its SMBH mass has reached to a certain level that has been found in simulations \citep{Zinger2020, Terrazas2020, Piotrowska2022}. 
We will come back to this in Section \ref{sec_QGBH}.

\begin{figure*}[htb]
    \centering
    \includegraphics[scale=0.29]{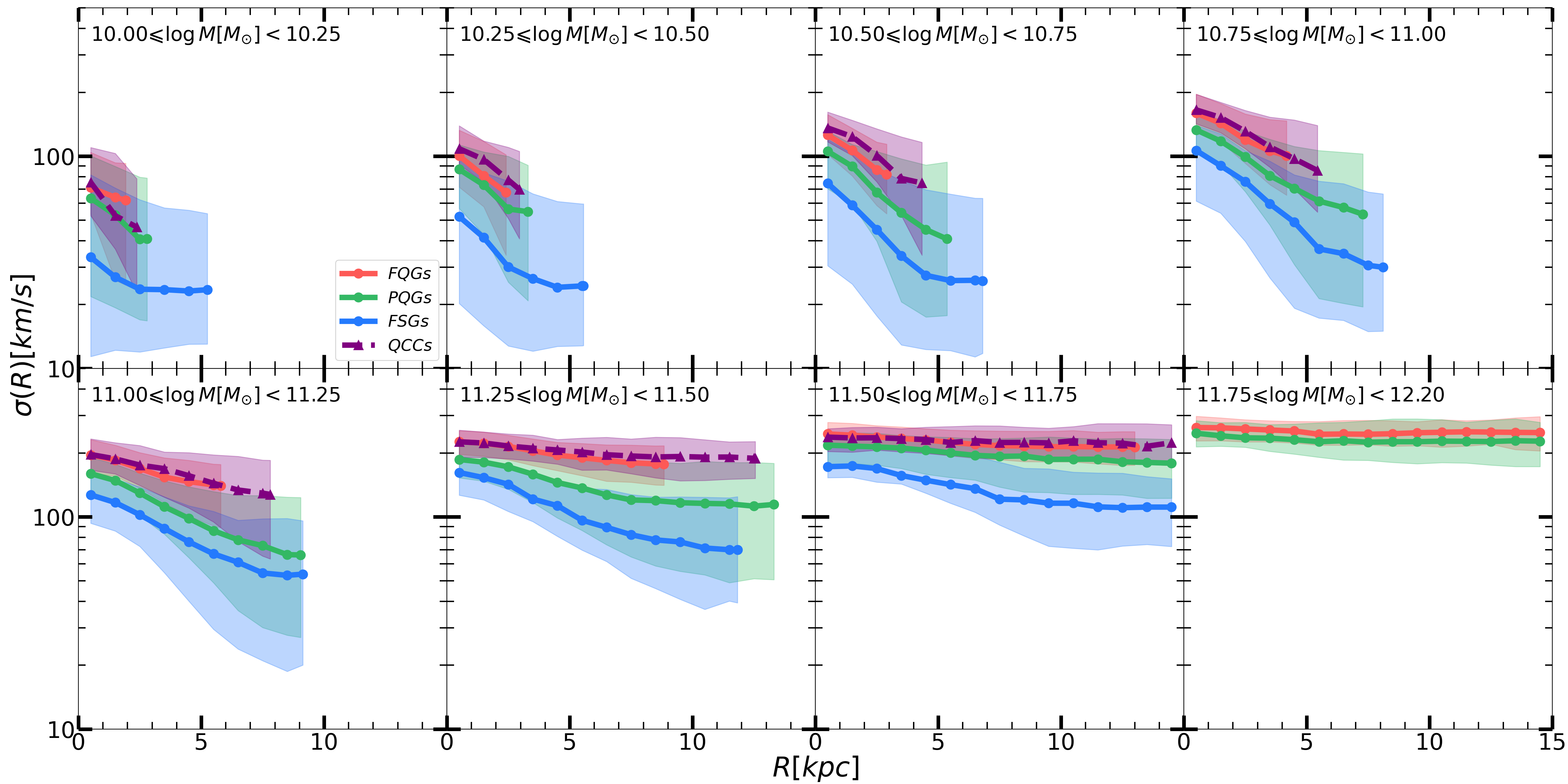}
    \caption{ Stellar velocity dispersion ($\sigma(R)$) profiles for FQGs (red), PQGs (green), FSGs (blue) and QCCs (purple). The solid and dashed lines show the median profiles of $\sigma(R)$ while the shaded areas are the 16th and 84th percentiles. 
     For galaxies, we show profiles up to the median of $1.5 R_{\rm e}$. For QCCs, we show the profiles to the median of $R_{\rm q}$. The results for galaxies and objects with different mass are presented in different panels. We ignore the results if 
     the number of galaxies or objects is less than 10.}\label{fig_sigma_prof}
\end{figure*}

\subsection{Quenched central cores}\label{sec_QCCs}

Recent IFU observations suggest that in may galaxies 
quenching of star formation is inside-out \citep[e.g.][]{LiC2015ApJ, WangE2018a}. 
This can also be seen in Figure \ref{fig_prof_fq}, which shows that PQGs usually have fully quenched cores in 
their centers while remain star-forming in outer parts. 
The ratio between the QCC mass, $M_{\rm q}$,  and $M_*$ 
increases with $\bar f_{\rm q}$, with the median 
$M_{\rm q}/M_*\approx 0.46$, 0.56 and 0.70 for 
$0.05\leq \bar f_{\rm q}<0.33$, $ 0.33\leq \bar f_{\rm q}< 0.66$, 
 $0.66\leq \bar f_{\rm q}\leq 0.95$, respectively. 
Figure \ref{fig_ssfr_mqms} shows the sSFR as a function of 
$M_{\rm q}/M_*$. At $\log\rm sSFR>-12$, the relation shows a clear 
negative trend that the sSFR of a galaxy decreases as the mass fraction of 
the QCC increases. At $\log\rm sSFR<-12$, no significant correlation 
can be seen between sSFR and $M_q/M_\star$. As discussed above and 
shown in Figure \ref{fig_ssfr-sigc}, PQGs with $\log\rm sSFR<-12$ have properties
similar to those of FQGs.  

We calculate the central velocity dispersion for QCCs using the same method 
as for galaxies. To do this, we need to know the radius, $R_{\rm e}$, 
for each QCC, which can be calculated from $R_{\rm q}$ as defined in Section \ref{sec_gclass}.  
Figure \ref{fig_mass_size} shows $R_{\rm q}$ versus $M_{\rm q}$.
For comparison, we also include the $R_{\rm e}$-$M_{*}$ relation for FQGs.
Both FQGs and QCCs follow a tight mass-size relation.
At the same mass, $R_{\rm q}$ of QCCs is larger than $R_{\rm e}$ of 
FQGs. This systematic difference is expected because $R_{\rm q}$ is 
defined in a way different from $R_{\rm e}$.
We perform linear regression for the two relations and the results are 
also shown in Figure \ref{fig_mass_size}. For FQGs, we have $\log R_{\rm FQG}(M_*)=0.79\log M_*/\Msun-8.20$. For QCCs, we have $\log R_{\rm QCC}(M_{\rm q})=0.76\log M_{\rm q}/\Msun-7.54$.
The effective radius, $R_{\rm e}$, of a QCC with $M_{\rm q}$ is defined as 
$R_{\rm e}(M_{\rm q})=R_{\rm q} R_{\rm FQG}(M_{\rm q})/R_{\rm QCC}(M_{\rm q})$. 
The underlying assumption in this definition is that QCCs have the same 
mass density profiles as FQGs of the same mass. Our examination shows 
that their surface mass density profiles are similar, and thus the assumption 
is valid. We then use the median $\sigma$ of spaxels within 0.2$R_{\rm e}$ 
to estimate the central velocity dispersion.
For simplicity, the new central velocity dispersion is also denoted 
as $\sigma_{\rm in}$. 

Figure \ref{fig_ms_sigc_QCCs} shows $\sigma_{\rm in}$ versus $M_{\rm q}$ for QCCs. 
For comparison, we also show $\sigma_{\rm in}$ versus $M_*$ for the corresponding 
PQGs in which QCCs are identified, and plot the QGSR to represent the the result 
for FQGs.
As expected, PQGs tend to lie below the QGSR, and the deviation from the QGSR 
increases with the decrease of $\bar f_{\rm q}$. In contrast, most data points for 
QCCs follow closely the QGSR, with a few exceptions to be discussed 
in more detail in Section \ref{sec_spe}. 
We can quantify this by estimating the `distances' of 
PQGs, QCCs and FQGs from the QGSR  using a parameter defined as    
\begin{equation}
    \chi^2=\frac{1}{N}\sum_{i=1}^N\frac{[\log{\sigma_{{\rm in},i}}-\log{\sigma_{\rm q}(M_{i})}]^2}{e_{\rm \sigma_{{\rm in},i}}^2+\alpha^2 e_{\rm M_{i}}^2+\Delta_{\rm q}^2},\\\label{eq_chi_QCC_PQG}
\end{equation}
where $\alpha$ and $\Delta_{\rm q}$ are the slope and the intrinsic scatter of the 
QGSR, $M_i$ is the $M_*$ or $M_{\rm q}$ of the $i$th galaxy or QCC, 
$\sigma_{{\rm in},i}$ is $\sigma_{{\rm in}}$ of the $i$th object,
and $N$ is the number of objects. For FQGs, we obtain 
$\chi^2_{\rm FQG}=1.03$. For galaxies with $0.66\leq\bar f_{\rm q}\leq0.95$, 
we obtain $\chi^2_{\rm PQG}=1.98$ and $\chi^2_{\rm QCC}=1.66$.   
For lower $\bar f_{\rm q}$, the difference between QCCs and PQGs is even 
larger: $\chi^2_{\rm PQG}=2.26$ and $\chi^2_{\rm QCC}=1.64$ for 
$0.33\leq\bar f_{\rm q}<0.66$; $\chi^2_{\rm PQG}=3.74$ and 
$\chi^2_{\rm QCC}=1.57$ for $0.05\leq\bar f_{\rm q}<0.33$. 
Note that the values of $\chi^2_{\rm QCC}$ are similar to 
$\chi^2_{\rm FQG}$, quite independent of  
$\bar f_{\rm q}$, while the values of $\chi^2_{\rm PQG}$ are significantly 
higher, especially for lower values of $\bar f_{\rm q}$. 
This demonstrates clearly that quenched central cores are similar to 
fully quenched galaxies in that they are both dynamical hot for their 
stellar mass. 

As mentioned in Section \ref{sec_gclass}, we require $R_{\rm q}\geq R_{\rm th}=4''$ 
to identify QCCs. The PSF FWHM of MaNGA is about $2''.5$, which is  
expected to smooth both the quenching and $\sigma$ profiles 
and to lead to some bias in the measurement.
This effect may be particularly significant for $M_{\rm q}$ and $R_{\rm q}$
but should be weak for $\sigma_{\rm in}$. This is because the quenching profile 
usually drops quickly around $R_{\rm q}$ (see Figure \ref{fig_prof_fq}), 
while $\sigma_{\rm in}$ is measured in the central region where the profile is 
typically flat (see Figure \ref{fig_sigma_prof}). 
Our tests using different radius threshold,  
$R_{\rm th}= 2''$, $3''$, $4''$ and $5''$, confirm the smoothing effect. 
In general, QCCs defined using different $R_{\rm th}$ all follow 
a similar $M_{\rm q}$-$\sigma_{\rm in}$ relation close to the 
QGSR. However, using a smaller $R_{\rm th}$ leads to larger scatter 
in the relation and, on average, gives slightly smaller $M_{\rm q}$ at 
given $\sigma_{\rm in}$. The bias becomes insignificant when $R_{\rm th}\geq4''$.
Note that QCCs can also be identified in FSGs of non-zero $\bar f_{\rm q}$ 
using smaller $R_{\rm th}$, and they also follow the QGSR. 
However, they are excluded from our analyses because they do not meet the 
requirement that $R_{\rm q}\geq R_{\rm th}=4''$.

In summary, the results presented in this section show that QCCs are 
similar to FQGs in that they are both dynamically hot 
(see also Section \ref{sec_2sig}). This suggests that the relation 
between star formation quenching and dynamical hotness exists not only 
for galaxies but also for regions inside them. In this sense,      
a FQG can be considered as a QCC. As we will show below, 
this relation also holds in outer regions of galaxies,
indicating that dynamical hotness is a necessary condition for 
quenching of star formation.

\begin{figure*}[htb]
    \centering
    \includegraphics[scale=0.29]{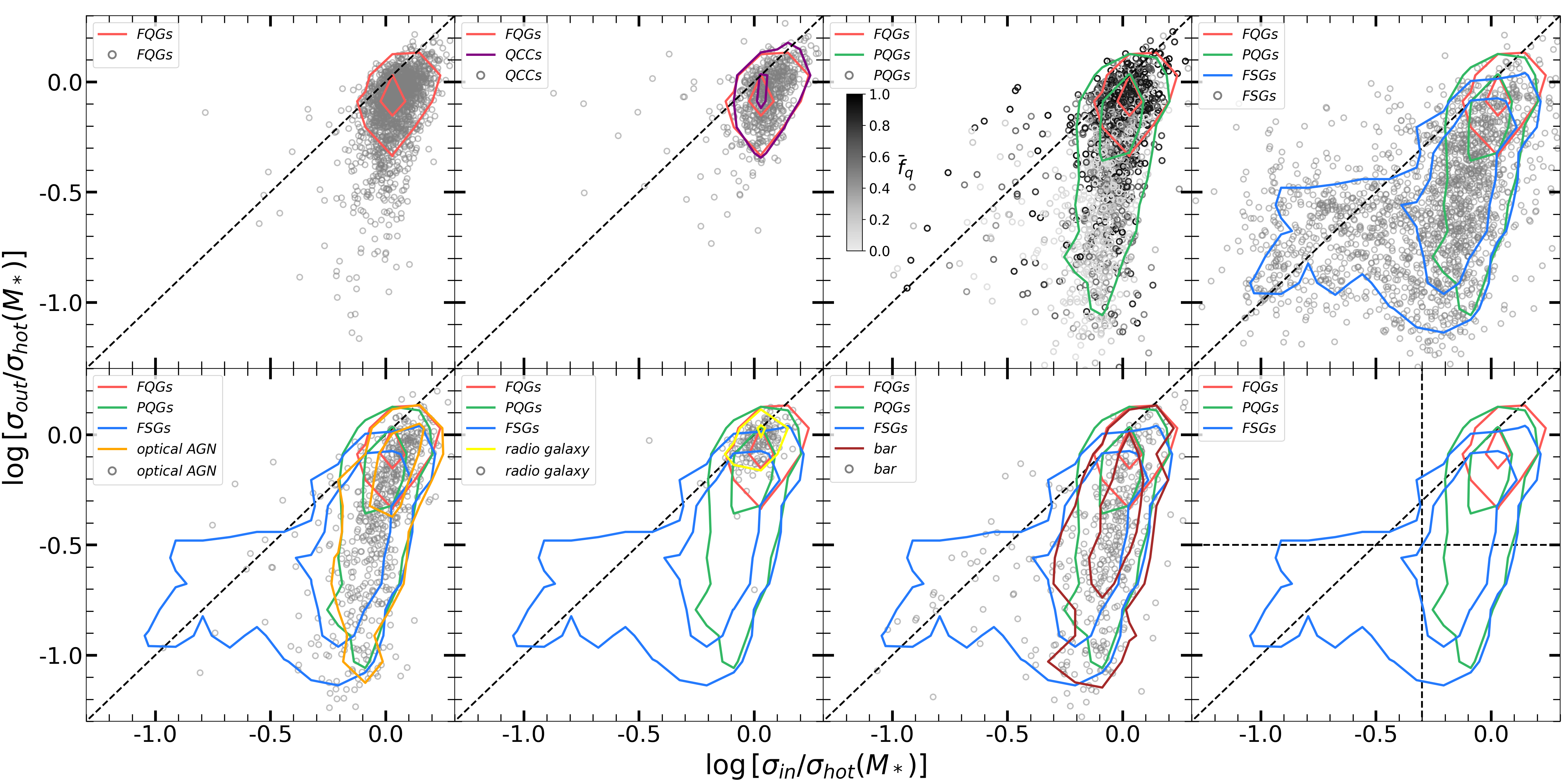}
    \caption{Two-$\sigma$ diagram: $\sigma_{\rm out}/\sigma_{\rm hot}(M_*)$ versus $\sigma_{\rm in}/\sigma_{\rm hot}(M_*)$. For QCCs, it is $\sigma_{\rm out}/\sigma_{\rm hot}(M_{\rm q})$ versus $\sigma_{\rm in}/\sigma_{\rm hot}(M_{\rm q})$. 
     The small circles show
     the results for FQGs (upper-left), QCCs (upper-middle-left), PQGs (upper-middle-right), FSGs (upper-right), optical AGN host galaxies (lower-left), radio galaxies (lower-middle-left) and barred galaxies (lower-middle-right). In the upper-middle-right panel, PQGs are shown with color-coded circles according to their $\bar{f}_{\rm q}$. In the lower-right panel, contour lines for FQGs (red), PQGs (green) and FSGs (blue) are shown. The inclined dashed lines show 1:1 relation. The horizontal and vertical dashed lines are the ones used to define horizontal and vertical sequences as discussed in Section \ref{sec_coe}.
     Contour lines with different colors correspond to different galaxies/objects as labeled in the corresponding panels. We show two contour levels, which enclose 40\% and 80\% of the galaxies/objects, respectively. 
     }
    \label{fig_2sig}
\end{figure*}

\begin{figure*}
    \centering
    \includegraphics[scale=0.45]{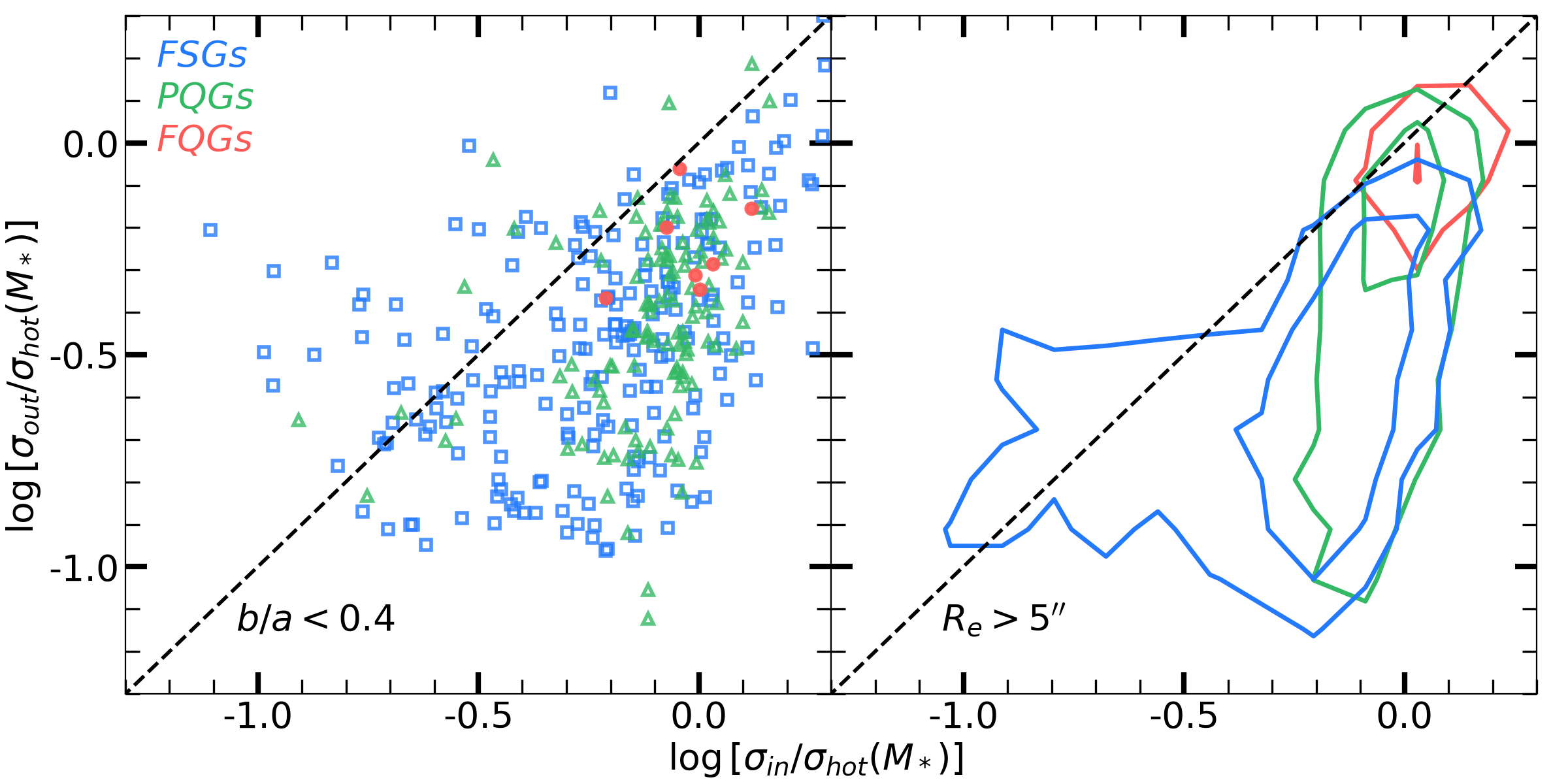}
    \caption{The left panel shows results for  galaxies with $b/a<0.4$. In the right panel, 
    $R_{\rm e}$ is required to be greater than $5''$. Contour levels correspond to 40\% and 80\%. FQGs, PQGs and FSGs are in red, green and blue, respectively.}\label{fig_2sig5}
\end{figure*}

\subsection{The Two-$\sigma$ diagram}\label{sec_2sig}

To better understand the connection between quenching and dynamical hotness,
we examine the velocity dispersion ($\sigma$) profile and its 
connection to quenching. Figure \ref{fig_sigma_prof} shows the median 
$\sigma$ profiles for different populations of galaxies and for QCCs. 
The shaded regions cover 16th to 84th percentile of the distribution of 
individual profiles. As one can see, FQGs and QCCs are always similar 
and have the highest velocity dispersion profiles; FSGs have the lowest 
velocity dispersion and PQGs lie in between.
For massive FQGs and QCCs, $\sigma$ is almost independent of 
radius. A mild decline with radius can be seen in lower-mass objects. 
In contrast,  FSGs,  in particular those of $\log M_*/\Msun<11$,  
exhibit a transition in their profile: the velocity dispersion first 
decreases gradually with radius and then remains at a roughly constant 
level of about $20-30\kms$ for low-mass galaxies and higher level 
for massive ones. All these indicate that dynamical hotness can vary 
within individual galaxies and the variation is systematically 
different for galaxies of different quenching properties.
The difference between different populations appears to be larger in 
outer regions, in contrast to the fact that the stellar mass surface density profile 
in the outer region is quite similar for different populations 
\citep[see][]{Fang2013ApJ}. This suggests that dynamical hotness  
may be more related to quenching than the local stellar mass surface density.

It is thus interesting to include the hotness in outer regions 
as an additional parameter to describe the dynamical properties
of galaxies. Here we use $\sigma_{\rm out}$ defined in Section \ref{sec_pipe3d}, 
and adopt its ratio with $\sigma_{\rm hot}$ to describe the outer 
hotness. In Figure \ref{fig_2sig}, we show the Two-$\sigma$ diagram, 
which is  $\sigma_{\rm out}/\sigma_{\rm hot}(M_*)$ versus 
$\sigma_{\rm in}/\sigma_{\rm hot}(M_*)$ for galaxies, and 
$\sigma_{\rm out}/\sigma_{\rm hot}(M_{\rm q})$ versus 
$\sigma_{\rm in}/\sigma_{\rm hot}(M_{\rm q})$ for QCCs.
As one can see, FQGs occupy a small region around 
$\log(\sigma_{\rm in}/\sigma_{\rm hot})\sim 0$ and 
$\log(\sigma_{\rm out}/\sigma_{\rm hot})\sim -0.05$ with size $0.1\sim0.2$ dex. 
Only a small fraction of FQGs have small $\sigma_{\rm out}/\sigma_{\rm hot}$. 
QCCs occupy almost the same region as FQGs, indicating 
that both QCCs and FQGs are dynamically hot in both inner and outer 
regions. PQGs are color-coded according to their $\bar{f}_{\rm q}$. They have a narrow distribution in
$\sigma_{\rm in}/\sigma_{\rm hot}$, with a median value slightly 
smaller than that of FQGs, but a much broad distribution in 
$\sigma_{\rm out}/\sigma_{\rm hot}$. They thus form a long vertical 
band extended to the region covered by FQGs. Along this vertical band with increasing $\sigma_{\rm out}/\sigma_{\rm hot}$, we can see an increasing trend of $\bar{f}_{\rm q}$ for PQGs accompanied by the growth of their QCCs. It strongly suggests that the spreading of dynamical hotness tightly correlates with the spreading of quenching. Figure \ref{fig_ssfr-sigc} shows that a small fraction of PQGs with
$\log\rm sSFR<-12$ have almost the same $\sigma_{\rm in}$ distribution as FQGs. These galaxies occupy the same region as FQGs in Two-$\sigma$ diagram. 
FSGs have two branches, horizontal one and vertical one. Galaxies in the horizontal branch 
are all relatively cold in outer regions, while their inner regions 
can either be dynamically cold or hot.  The vertical branch
overlaps with the PQG band and also extends to the region covered by FQGs, 

Comparing the Two-$\sigma$ diagram with Figure \ref{fig_sigc-ms},
we find that a significant fraction of PQGs and FSGs have 
smaller $\sigma_{\rm out}$ than FQGs, and so the Two-$\sigma$ diagram 
can be used to separate different populations of galaxies more effectively 
than the $M_*$-$\sigma_{\rm in}$ diagram. 
For example, there are 1,073 FQGs, 484 PQGs and 130 FSGs enclosed by the 
circle of a radius of 0.15 dex, centered at (0,-0.05). 
Among the 484 PQGs, 255 of them have $\log\rm sSFR<-12$. We find that about 71\% of 
dynamically hot galaxies have $\log\rm sSFR<-12$ and 88\% of hot galaxies have $\log\rm sSFR<-11$.
This clearly indicates a potential  
link between dynamical hotness and star formation quenching.
In what follows, we use the Two-$\sigma$ diagram to define dynamic 
hotness of a whole object: an object is said to be dynamically hot when 
both $\sigma_{\rm in}$ and $\sigma_{\rm out}$ are close to 
the value of $\sigma_{\rm hot}$ corresponding to its stellar mass.
Thus, both FQGs and QCCs are dynamically hot. 
Note that a small fraction of FSGs are also dynamically hot. 
This seems to suggest that dynamical hotness is a necessary, 
but not a sufficient condition for quenching. As we will   
discuss later, the existence of these hot star-forming objects  
suggests a potential path for galaxies to quench star formation.  

The vertical band and horizontal branch represent the other two major 
dynamical status. The horizontal branch is dominated by FSGs, and only 
a few PQGs and FQGs are found in this branch. Galaxies on this branch have 
cold dynamics in both the inner and outer regions. The vertical band  
consists mainly of PQGs and FSGs. Galaxies in this band
have relatively hot stellar component in the inner region but 
cold in the outskirts. The two dynamical status are roughly separated 
at $\log(\sigma_{\rm in}/\sigma_{\rm hot})\sim-0.3$. 
As we will discuss in Section \ref{sec_coe}, the existence of well-defined branches 
of galaxy distribution in the Two-$\sigma$ diagram provides important clues
about the coevolution between star formation in galaxies and their dynamical 
structure.

We also show the distributions of optically-selected AGNs, radio galaxies and 
barred galaxies (Section \ref{sec_sample}) in the lower panels of Figure \ref{fig_2sig}. 
These three types of galaxies all have their specific regions of residence 
in the Two-$\sigma$ diagram. Both barred galaxies and optical AGNs reside
on the vertical branch where PQGs are found, while radio galaxies can only 
be found in the top-right corner of the Two-$\sigma$ diagram where FQGs reside. 
All of these suggest a potential link among star formation quenching, AGN activities 
and feedback, and galaxy morphology and dynamical structure. We will
come back to this issue in Section \ref{sec_coe}.

Measurements of velocity dispersion may be affected by rotation
in edge-on galaxies. To check whether or not this has a significant 
impact on our results, we select galaxies with axis ratios 
$b/a<0.4$ and show them in the Two-$\sigma$ diagram plotted 
in Figure \ref{fig_2sig5}. In the upper-left corner 
($\log(\sigma_{\rm out}/\sigma_{\rm hot})>-0.5$ and $\log(\sigma_{\rm in}/\sigma_{\rm hot})<-0.4$), 
about 19\% of FSGs have $b/a<0.4$, significantly higher than the fraction 
for the total FSGs (12\%). For PQGs, the fraction of edge-on galaxies is 
about 8\% for the total population and about 14\% in the upper-left 
corner. This suggests that the velocity dispersion of some galaxies in the 
upper-left corner is systematically overestimated because the rotation motion 
is not well subtracted.  However, it is worthwhile noting that the majority of 
edge-on galaxies overlap with the main population in the Two-$\sigma$ diagram. 
This indicates that random motion and rotation can be well separated 
for most of edge-on galaxies.

If the angular size of a galaxy is comparable to
the spaxel size (0''.5), the measurement of $\sigma$ may also be affected 
by rotation. To check this, we re-plot the Two-$\sigma$ diagram using galaxies 
with $R_{\rm e}>5''$ in the right panel of Figure \ref{fig_2sig5}. The general 
trend does not change, but a significant fraction of FSGs around the dynamically hot systems now disappear. This indicates that the velocity dispersion of
some of these FSGs is overestimated. Inspections of the 
images of these galaxies showed that some are indeed edge-on. However, adopting 
a cut in angular size can lead to severe bias. As shown in Section \ref{sec_coe}, 
FSGs with high $\sigma_{\rm in}$ and $\sigma_{\rm out}$ tend to have small physical sizes, 
and adopting a cut in 
the angular size will eliminate many of these compact galaxies, 
which may lead to bias. We thus do not apply such a cut in our analyses.

\begin{figure*}[htb]
    \centering
    \includegraphics[scale=0.4]{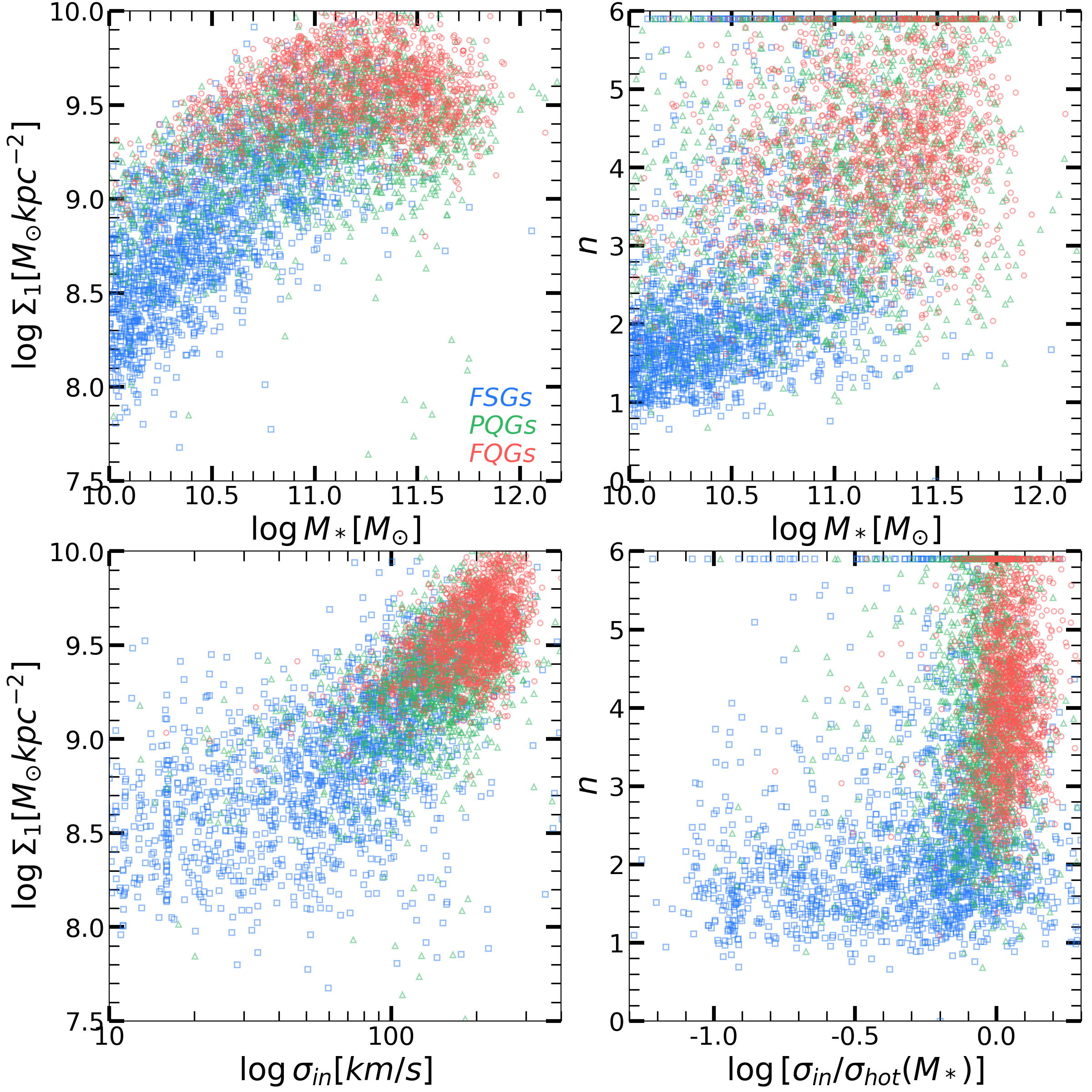}
    \caption{Top left panel: $\Sigma_1$ versus stellar mass; top right panel: S$\rm \Acute{e}$rsic index $n$ (r band) versus stellar mass; bottom left panel: $\Sigma_1$ versus $\sigma_{\rm in}$; bottom right panel: S$\rm \Acute{e}$rsic index $n$ versus $\sigma_{\rm in}/\sigma_{\rm hot}(M_*)$. FSGs, PQGs and FQGs are shown in blue squares, green triangles and red circles, respectively. }
    \label{fig_Sig1_sersicn_sigc_ms}
\end{figure*}

\subsection{Other structural parameters}\label{sec_otstr}

In the literature, many parameters have been used to characterize galaxy structure. 
These parameters can also be used to understand the connection between galaxy 
structure and star formation quenching. In this section, we present analyses 
similar to those shown in Section \ref{sec_dhqs} for two additional structural 
parameters, $\Sigma_1$ and S$\rm \Acute{e}$rsic index $n$, as defined  
and measured in Section \ref{sec_pipe3d} and Section \ref{sec_sample}. We have also checked other parameters, such as 
the bulge-to-total mass ratio and profile concentration. The results are broadly 
consistent with those based on $\sigma_{\rm in}$, $\Sigma_1$ and $n$, and are
not shown here for brevity.

The first panel of Figure \ref{fig_Sig1_sersicn_sigc_ms} shows 
$\Sigma_1$ versus $M_*$ for FQGs, PQGs and FSGs. Since QCCs are similar to 
FQGs, they are not shown here for clarity. In general, $\Sigma_1$ increases with 
$M_*$. At given $M_*$, the surface density is the highest for FQGs and lowest 
for FSGs. These results are broadly consistent with those shown in \cite{Fang2013ApJ} 
and \cite{Barro2017ApJ}. We find that the increasing trend appears to be reversed at 
$\log M_*/\Msun>$11.5, consistent with previous results \citep[e.g.][]{Graham2003}.
This happens because the most massive elliptical galaxies have flattened 
cores. The quenched galaxies in \cite{Fang2013ApJ} and \cite{Barro2017ApJ} 
exhibit a stronger correlation between $M_*$ and $\Sigma_1$ than our 
FQGs because of different selection criteria used; they only 
considered galaxies with $\log M_*/\Msun<$11.5. We also show the relation
between $\Sigma_1$ and $\sigma_{\rm in}$ in the lower left panel.
Consistent with \cite{Fang2013ApJ}, the two quantities are tightly correlated 
with each other, and the correlation becomes weaker as $\sigma_{\rm in}$ 
decreases.

As shown in \cite{Barro2017ApJ}, galaxy distribution in the $\Sigma_{1}$-sSFR plane
also exhibit an L-shaped structure, suggesting that $\Sigma_{1}$ is 
also a structural parameter that is closely related to star formation quenching.
The reasons for us to adopt $\sigma_{\rm in}$ instead of $\Sigma_{1}$ are the following. 
First, the L-shape structure in the $\Sigma_{1}$-sSFR diagram is looser, as 
shown by the low-redshift result in figure 6 of \citet{Barro2017ApJ}, than that 
in the $\sigma_{\rm in}$-sSFR diagram. 
Second, the correlation between $\Sigma_1$ and $M_*$ reverses at the massive end, 
as shown in Figure \ref{fig_Sig1_sersicn_sigc_ms}, which complicates the 
interpretation of the $\Sigma_1$-sSFR relation. Third, previous studies have 
shown that $\sigma_{\rm in}$ is more closely related to galaxy quenching 
than other structural parameters \citep[e.g.][]{Wake2012, Bluck2016, Bluck2020, Brownson2022}. 
Fourth, as shown in Sections \ref{sec_2sig} and \ref{sec_coe}, the velocity 
dispersion can be used to separate galaxies of different dynamical status.
Finally, it is more straightforward to connect velocity dispersion to supermassive black 
hole mass that is thought to be closely related to galaxy quenching (see Section \ref{sec_QGBH}).

Next, let us examine using the S$\rm \Acute{e}$rsic index $n$ as a structural parameter. 
This index is often used to describe the galaxy morphology and surface density profile. 
A galaxy with $n=1$ is a pure exponential disk, while $n\geq 4$ is usually found  
for elliptical galaxies. The S$\rm \Acute{e}$rsic indices used here were derived 
from the SDSS $r$-band images, taken from the NYU-VAGC catalog \citep{Blanton2005}. 
An upper limit of $n=6$ was set in the fitting. The upper right panel 
of Figure \ref{fig_Sig1_sersicn_sigc_ms} shows $n$ versus $M_*$. 
There is a loose correlation between the two quantities, with more massive galaxies 
tending to have higher $n$. Moreover, FQGs on average have higher $n$ than FSGs 
at given $M_*$. We also present $n$ versus $\sigma_{\rm in}/\sigma_{\rm hot}$
in the lower right panel. Here one can see an L-shaped structure in the 
distribution. Most galaxies at $\log(\sigma_{\rm in}/\sigma_{\rm hot})<-0.3$ have $n$ between 1 and 2, 
independent of  $\sigma_{\rm in}/\sigma_{\rm hot}$. As
$\log(\sigma_{\rm in}/\sigma_{\rm hot})>-0.3$, the value of $n$ 
ranges from 1 to 6. FQGs have a very broad $n$ distribution and 
are mixed with other galaxies, indicating that it is difficult to 
use $n$ to separate galaxies of different quenching status. 
We have also checked the relation between sSFR and $n$, and found a 
negative and loose correlation. No L-shaped structure is seen 
in the distribution of galaxies in the sSFR-$n$ plane.

\section{Co-evolution of dynamical status and star formation}\label{sec_coe}

The Two-$\sigma$ diagram shows that galaxy distribution 
in this diagram is L-shaped. 
Here we investigate in more detail galaxy 
properties in this distribution by separating galaxies into two 
dynamical sequences (see the lower-right panel of Figure \ref{fig_2sig}):
\begin{itemize}
\item \textit{Horizontal sequence:} galaxies with 
         $\log(\sigma_{\rm out}/\sigma_{\rm hot})<-0.5$;
\item \textit{Vertical sequence:} galaxies with 
         $\log(\sigma_{\rm in}/\sigma_{\rm hot})>-0.3$.
\end{itemize}
This separation roughly separates the two main components seen in the 
Two-$\sigma$ diagram. We note that the two sequences overlap in the 
lower-right corner. 
In the following, we investigate how galaxy 
structure and activity change along the two sequences. 
We will also examine galaxies outside the main sequences.
The goal here is to understand how the dynamical and structural
properties, star formation and nucleus activities are connected.

\begin{figure*}[h]
    \centering
    \includegraphics[scale=0.39]{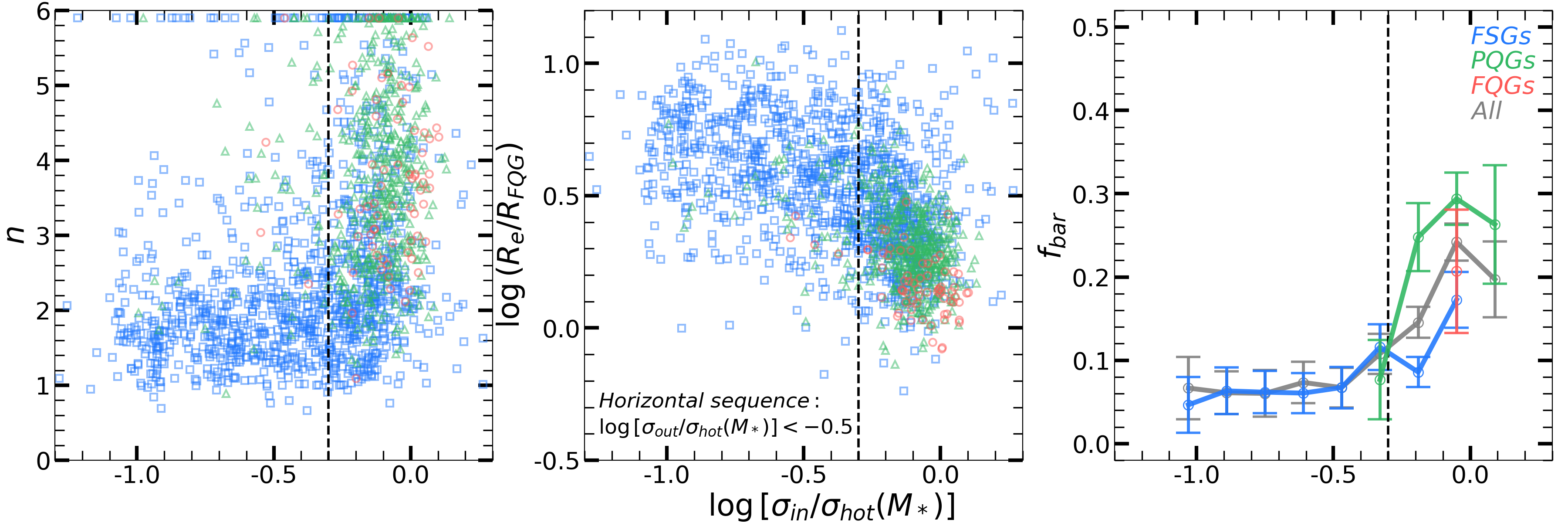}
    \caption{Galaxy structure along horizontal sequence ($\log(\sigma_{\rm out}/\sigma_{\rm hot})<-0.5$). FSGs, PQGs and FQGs are shown in blue, green and red, respectively. Left panel: S$\rm \Acute{e}$rsic index $n$ as a function of $\log(\sigma_{\rm in}/\sigma_{\rm hot})$. Middle panel: $\log[R_{\rm e}/R_{\rm FQG}(M_*)]$ as a function of $\log(\sigma_{\rm in}/\sigma_{\rm hot})$, where $R_{\rm FQG}(M_*)$ is the best fitting for the mass-size relation of FQGs shown in Figure \ref{fig_mass_size}. Right panel: Barred galaxy fraction ($f_{\rm bar}$) as a function of $\log(\sigma_{\rm in}/\sigma_{\rm hot})$. Gray line shows the results for all galaxies and error bars are obtained with bootstrap method. The vertical dashed lines show $\log(\sigma_{\rm in}/\sigma_{\rm hot})=-0.3$. We do not show the results in bins with galaxy number 
    less than 20. }\label{fig_hs_morph}
\end{figure*}

\subsection{The horizontal sequence}\label{sec_hs}
 
Here we examine three structural parameters,  S$\rm \Acute{e}$rsic index $n$,
galaxy size ($R_{\rm e}$) and the fraction of barred galaxies ($f_{\rm bar}$). 
The left panel of Figure \ref{fig_hs_morph} shows $n$ as a function of 
$\sigma_{\rm in}/\sigma_{\rm hot}$. We can see that most of the 
galaxies with $\log(\sigma_{\rm in}/\sigma_{\rm hot})<-0.3$ 
have $n$ between 1 and 2, quite independent of $\sigma_{\rm in}/\sigma_{\rm hot}$. 
Our visual inspections showed that most of these galaxies possess 
spiral arms, indicating that the horizontal sequence at 
$\log(\sigma_{\rm in}/\sigma_{\rm hot})<-0.3$ is dominated by disk galaxies. 
This is consistent with the small velocity dispersion observed in both inner and outer 
regions. At $\log(\sigma_{\rm in}/\sigma_{\rm hot})>-0.3$, the 
value of $n$ suddenly becomes much larger and has a much broader 
distribution, indicating that galaxies become more concentrated and diverse.
The data thus suggests a morphological transition 
at $\log(\sigma_{\rm in}/\sigma_{\rm hot})\sim -0.3$. 
Such a transition can also be seen in the other two structural parameters. 
The middle panel shows galaxy size ($R_{\rm e}$) as a function of 
$\sigma_{\rm in}/\sigma_{\rm hot}$. Here galaxy sizes are normalized by 
the best-fit $M_*$-$R_{\rm e}$ relation for FQGs (see Figure \ref{fig_mass_size}). 
At $\log(\sigma_{\rm in}/\sigma_{\rm hot})<-0.3$, galaxy size is almost independent of
$\sigma_{\rm in}/\sigma_{\rm hot}$, while at 
$\log(\sigma_{\rm in}/\sigma_{\rm hot})>-0.3$ there is a sudden drop. 
This again suggests a transition to higher concentration, similar to that 
seen in $n$. The barred galaxy fraction, $f_{\rm bar}$, is shown as a function of 
$\sigma_{\rm in}/\sigma_{\rm hot}$ in the right panel. 
At $\log(\sigma_{\rm in}/\sigma_{\rm hot})<-0.3$, $f_{\rm bar}$ is almost a 
constant at a low value around 6\%. But the fraction increases quickly at 
$\log(\sigma_{\rm in}/\sigma_{\rm hot})>-0.3$, in particular for PQGs.
This suggests that the presence of a bar is somehow linked to 
the dynamical hotness in the inner region of a galaxy (see also Figure \ref{fig_2sig}).  


\begin{figure*}
    \centering
    \includegraphics[scale=0.45]{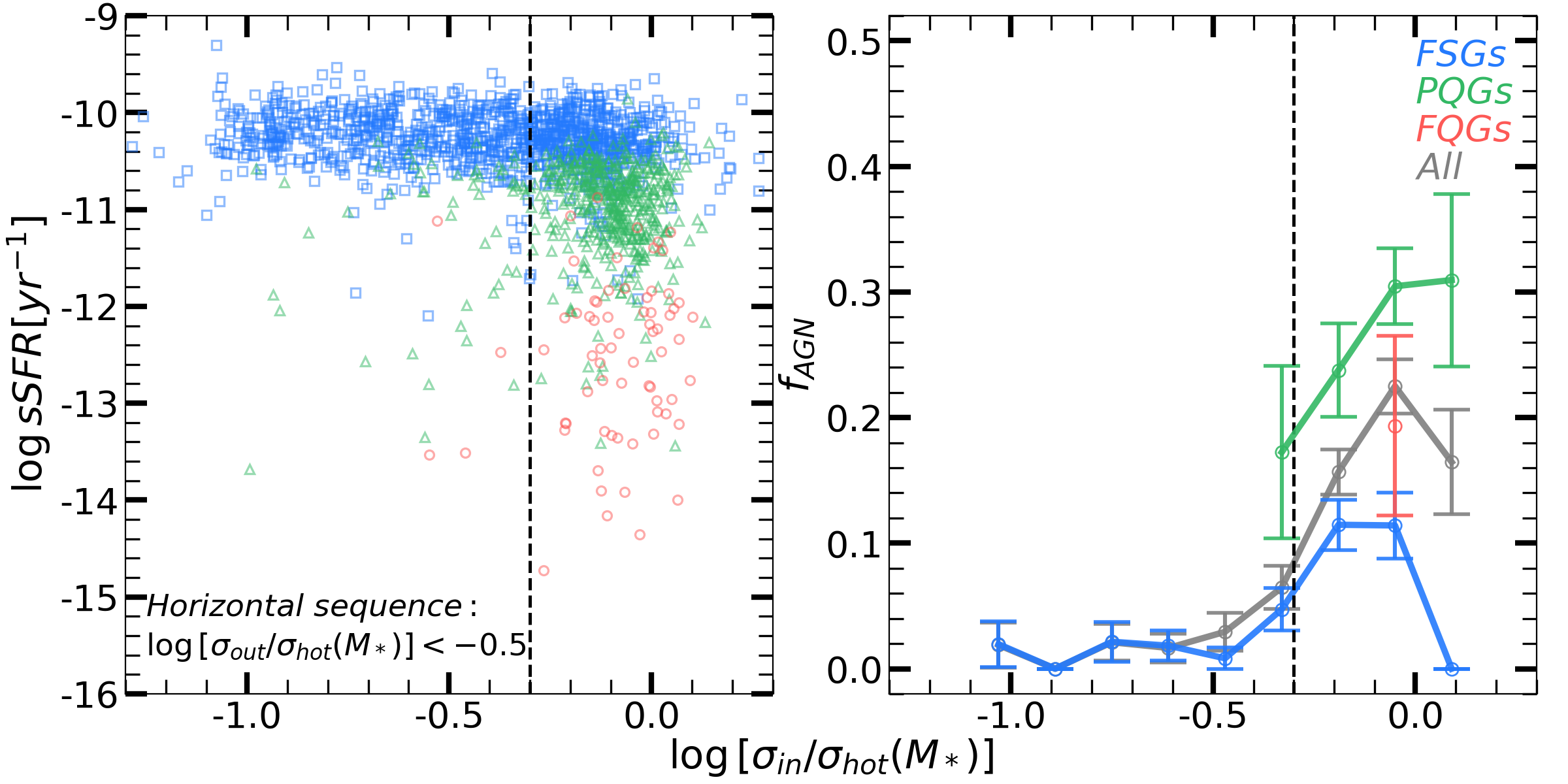}
    \caption{Galaxy activity along the horizontal sequence ($\log(\sigma_{\rm out}/\sigma_{\rm hot})<-0.5$). FSGs, PQGs and FQGs are shown in blue, green and red, respectively. Left panel: $\rm sSFR$ as a function of $\log(\sigma_{\rm in}/\sigma_{\rm hot})$.  Right panel: Similar to the right panel of Figure \ref{fig_hs_morph} but for optical AGN fraction ($f_{\rm AGN}$) as a function of $\log(\sigma_{\rm in}/\sigma_{\rm hot})$. The vertical dashed lines show $\log(\sigma_{\rm in}/\sigma_{\rm hot})=-0.3$.}\label{fig_hs_activity}
\end{figure*}

In Figure \ref{fig_hs_activity}, we show the sSFR as a function of 
$\sigma_{\rm in}/\sigma_{\rm hot}$. An L-shaped distribution, similar to that 
in Figure \ref{fig_ssfr-sigc}, can be clearly seen. 
At $\log(\sigma_{\rm in}/\sigma_{\rm hot})<-0.3$, galaxies have 
$\log\rm sSFR\sim -10.2$, independent of velocity dispersion. 
A quick drop in the sSFR occurs around $\log(\sigma_{\rm in}/\sigma_{\rm hot})\sim-0.3$, 
and a large number of PQGs emerge above this value of 
$\log(\sigma_{\rm in}/\sigma_{\rm hot})$. This transition 
clearly indicates that quenching of central parts of galaxies occurs 
when the central parts become dynamically hot, although the whole galaxies 
are dynamically cold and star forming, as discussed in Section \ref{sec_QCCs}.

AGN feedback is thought to be one of the major mechanisms to quench star formation.
It is thus interesting to examine the change of AGN activities along the horizontal 
sequence. The right panel of Figure \ref{fig_hs_activity} shows the fraction 
of optically selected AGNs, $f_{\rm AGN}$, as a function of 
$\sigma_{\rm in}/\sigma_{\rm hot}$. Here again, we see a dramatic change in
$f_{\rm AGN}$ at $\log(\sigma_{\rm in}/\sigma_{\rm hot})\sim-0.3$. 
The fraction is very low, usually less than 2\%, at 
$\log(\sigma_{\rm in}/\sigma_{\rm hot})<-0.3$. 
At $\log(\sigma_{\rm in}/\sigma_{\rm hot})>-0.3$, 
the fraction can reach as high as 23\% for the whole galaxy population. 
The jump is particularly large for PQGs, from zero to about 30\%. 
The result is consistent with that shown in Figure \ref{fig_2sig}.
Given that the AGN duty cycle is typically $10^8$ years 
\citep[e.g.][]{Marconi2004}, the high AGN fraction suggests that almost 
all of the galaxies at $\log(\sigma_{\rm in}/\sigma_{\rm hot})>-0.3$
have gone through AGN phases in a few hundred-million years in the past. 

To summarize, our results show that galaxy structure and activity 
make a fast transition along the horizontal sequence when the central velocity dispersion 
reaches about half of the velocity dispersion of a dynamically hot system 
according to their mass: galaxies become more concentrated, morphologically 
more diverse, with stronger bar and AGN activities as they cross the 
transition line. This transition also marks the onset of star-forming 
quenching in central regions of galaxies. All these indicate a close 
link of dynamical hotness with structural properties of galaxies, 
AGN feeding and star formation quenching. We will come back to this in 
Section \ref{sec_track}.

\subsection{The vertical sequence}\label{sec_vs}

\begin{figure*}
    \centering
    \includegraphics[scale=0.39]{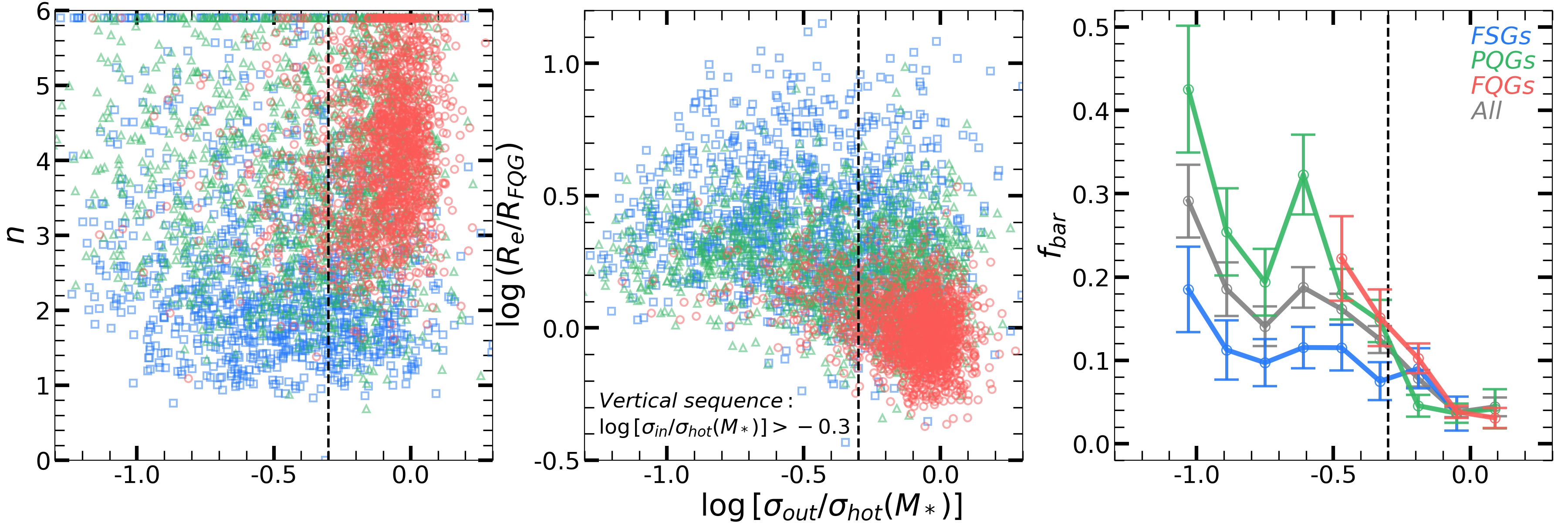}
    \caption{Similar to Figure \ref{fig_hs_morph} but here we show the structural parameters for galaxies along the vertical sequence ($\log(\sigma_{\rm in}/\sigma_{\rm hot})>-0.3$) as functions of $\log(\sigma_{\rm out}/\sigma_{\rm hot})$. The vertical dashed lines show $\log{\sigma_{\rm out}/\sigma_{\rm hot}}=-0.3$.}\label{fig_vs_morph}
\end{figure*}

\begin{figure*}
    \centering
    \includegraphics[scale=0.45]{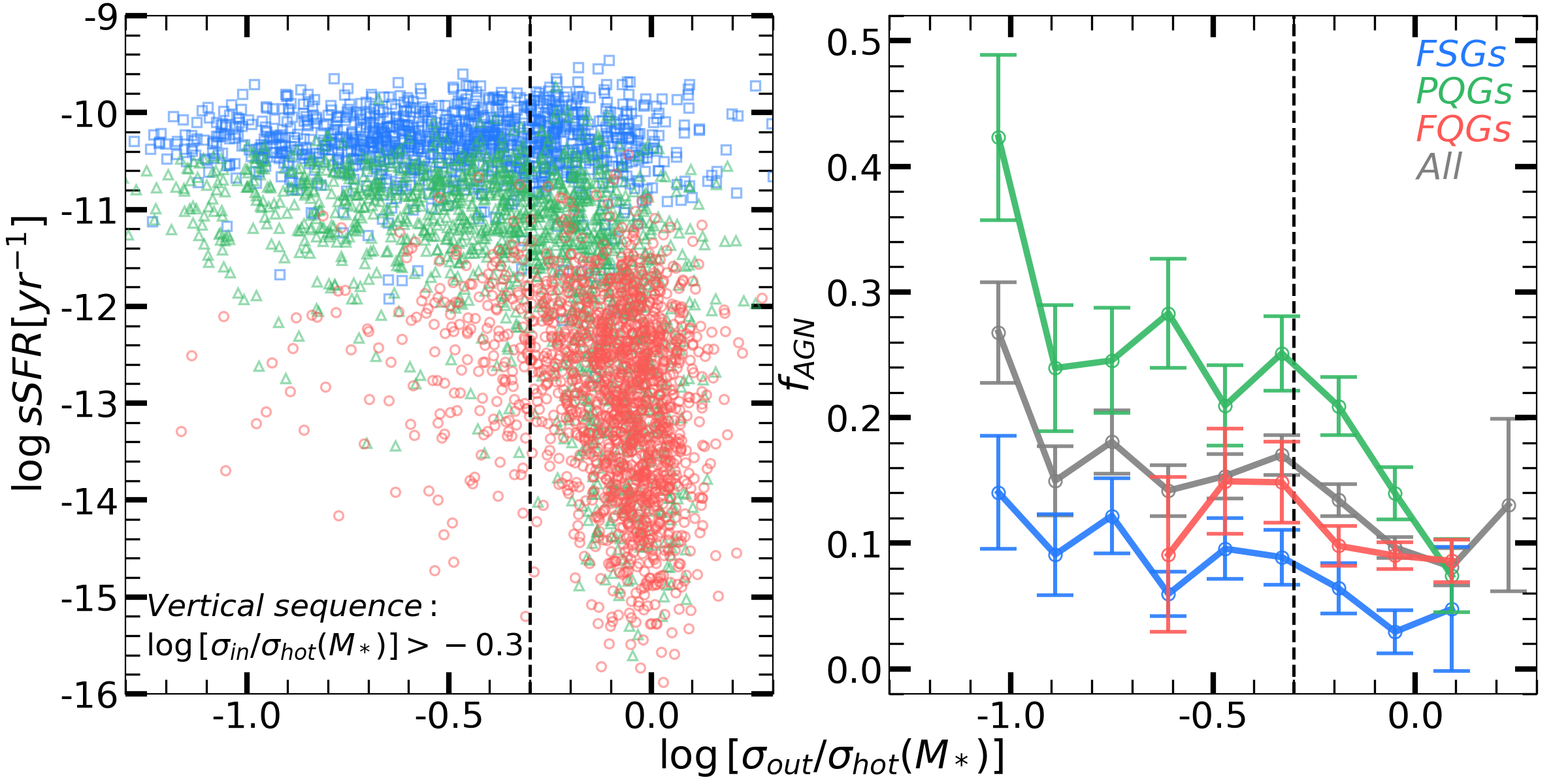}
    \caption{Similar to Figure \ref{fig_hs_activity} but here we show sSFR and $f_{\rm AGN}$ for galaxies along the vertical sequence ($\log(\sigma_{\rm in}/\sigma_{\rm hot})>-0.3$) as functions of $\log(\sigma_{\rm out}/\sigma_{\rm hot})$. The vertical dashed lines show $\log(\sigma_{\rm out}/\sigma_{\rm hot})=-0.3$.}\label{fig_vs_activity}
\end{figure*}


The left panel of Figure \ref{fig_vs_morph}
shows the S$\rm \Acute{e}$rsic index, $n$, as a function of 
$\sigma_{\rm out}/\sigma_{\rm hot}$ along the vertical sequence 
defined above. In general, all the three populations of galaxies, 
FSGs, PQGs and FQGs, have very broad distributions in $n$, 
indicating that galaxies in this sequence have diverse morphology.  
For a given population, there is no significant dependence of $n$ on 
$\sigma_{\rm out}/\sigma_{\rm hot}$, although different populations 
are distributed differently. The middle panel shows 
$R_{\rm e}/R_{\rm FQG}(M_*)$ as a function of 
$\sigma_{\rm out}/\sigma_{\rm hot}$.  
The size ratio decreases with increasing $\sigma_{\rm out}/\sigma_{\rm hot}$, 
and the trends for PQGs and FSGs are similar. 
The right panel shows that the fraction of barred galaxies gradually 
declines from 30\% to 4\% as $\sigma_{\rm out}/\sigma_{\rm hot}$
increases from $\sim 0.1$ to $\sim 1$, and the trend exists for all the three 
populations, being the strongest for PQGs. Combining these results with those  
for the horizontal sequence, we can conclude that galaxy bars prefer to live 
in galaxies located in the lower-right corner of the Two-$\sigma$ diagram, where 
a galaxy is dynamically hot in the central region but cold in the outskirt.

Figure \ref{fig_vs_morph} shows that a fraction of FSGs have sizes 
larger than those of PQGs of the same mass. To understand the difference, 
we selected FSGs with $\log(R_{\rm e}/R_{\rm FQG})>0.6$ and checked their 
properties. Most of these galaxies have $n$ between 1 and 2, suggesting that
they are spiral galaxies. In comparison, PQGs usually have 
$n>2$ (see the left panel). In the Two-$\sigma$ diagram, FSGs tend 
to be located closer to the limit of the vertical sequence
at $\log(\sigma_{\rm in}/\sigma_{\rm hot})=-0.3$ than PQGs. 
This suggests that the difference between FSGs and PQGs may be mainly in  
their $\sigma_{\rm in}$ distribution.

The left panel of Figure \ref{fig_vs_activity} shows the sSFR 
versus $\sigma_{\rm out}/\sigma_{\rm hot}$. FSGs have almost 
constant sSFR, independent of $\sigma_{\rm out}/\sigma_{\rm hot}$, 
similar to that along the horizontal sequence. The independence 
of the lower bound of sSFR is caused by the D4000 threshold used to 
define FSGs, but the presence of the roughly constant upper 
bound in sSFR indicates that there is an upper limit of sSFR 
that is quite independent of dynamical hotness.  
The sSFR of PQGs appears to be lower than that of FSGs,
as expected from their definition, and decreases gradually with 
increasing $\sigma_{\rm out}/\sigma_{\rm hot}$ at 
$\log(\sigma_{\rm out}/\sigma_{\rm hot})<-0.3$. It is consistent with the increasing
trend of $\bar f(q)$ with increasing $\sigma_{\rm out}/\sigma_{\rm hot}$ shown in Figure \ref{fig_2sig}.
At $\log(\sigma_{\rm out}/\sigma_{\rm hot})\sim-0.3$, however,   
the sSFR drops quickly and the FQG population emerges.  
The overall distribution of galaxies is L-shaped. 
This demonstrates that the L-shaped distribution shown in Figure \ref{fig_ssfr-sigc} 
contains two transitions. The first occurs when the central region 
of a galaxy becomes dynamically hot, i.e. when 
$\log(\sigma_{\rm in}/\sigma_{\rm hot})$ reaches $-0.3$ 
(Figure \ref{fig_hs_activity}). This corresponds to the emergence of PQGs 
and QCCs. The second transition occurs when a galaxy becomes dynamically hot 
over its entire body, i.e. when $\log(\sigma_{\rm out}/\sigma_{\rm hot})$
reaches $\sim-0.3$ (Figure \ref{fig_vs_activity}). This corresponds to the 
emergence of the FQG population. Note, however, when considered 
as systems of their own mass, QCCs occupy the same region as FQCs, 
which implies that the presence of a dynamically cold component 
in the outer part of a galaxy does not have a strong impact on the 
inner hot component. 

Optical AGNs reside throughout the vertical sequence (Figure \ref{fig_2sig}). 
We show the fraction of optical AGNs along the vertical sequence in the right panel of Figure 
\ref{fig_vs_activity}. The fraction is the highest for PQGs and 
the lowest for FSGs. The optical AGN fraction declines  
as $\sigma_{\rm out}/\sigma_{\rm hot}$ increases, and the decline 
becomes stronger at $\log(\sigma_{\rm out}/\sigma_{\rm hot})>-0.3$. 
Combining these results with the corresponding results for the horizontal 
sequence, we can see that optical AGNs are preferentially found in galaxies 
located in the lower-right corner of the Two-$\sigma$ diagram. 
This is similar to the fraction of barred galaxies, suggesting 
a potential connection between these two populations of galaxies.
Radio galaxies, which are also referred to as AGNs, are all dynamically hot,
which is consistent with their hosts being elliptical galaxies.
Most (151/156) of our radio galaxies are classified as low-excitation 
radio mode and most of them have $\log M_*/\Msun>11$,  
indicating that they might be powered by Bondi accretion of hot gas. 
Almost no radio galaxy is found to be dynamically hot in 
the central region and cold in the outskirt, indicating 
the presence of a radio AGN is very sensitive to the dynamical 
state even in the outer part of its host.

Images of dynamically hot galaxies may provide additional information about 
underlying processes relevant to the observed connection of galaxy properties.  
We thus selected galaxies within a circle of radius of 0.15 dex, 
centered at (0,-0.05), in the Two-$\sigma$ diagram.
Most of the FQGs and PQGs in this region appear to be round 
and smooth, consistent with being dynamically hot.
FSGs in this region usually appear to be compact, with sizes comparable to 
those of the FQGs of the same mass. We inspected the SDSS images 
of these galaxies and found that some FSGs appear to be 
edge-on spirals. However, the fraction of FSGs with $b/a<0.4$ is only 
about 12\%, comparable to the FSG population located outside 
the region. About 20\% of the FSGs in this region show spiral arms,  
and most of the spiral arms appear asymmetric.
More interestingly, more than half of FSGs in this region
show significant asymmetric, sometimes clumpy, structures.
These hot FSGs thus appear to have experienced significant 
interactions and/or mergers in the recent past. 

A small fraction of galaxies in the vertical sequence have very large 
central velocity dispersion (Figure \ref{fig_2sig}).
We selected galaxies with $\log\sigma_{\rm in}\geq\log\sigma_{\rm q}(M_*)+3\Delta_{\rm q}$, 
which gave 3 FQGs, 6 PQGs and 22 FSGs. 
We inspected the 22 images of FSGs and found that only four are regular, 
edge-on disks. The large $\sigma_{\rm in}$ for these disk galaxies 
is caused by rotation. The rest FSGs all exhibit clear signature for 
galaxy interaction and/or merger, showing close companions, 
asymmetric structure and tidal tails. The violent interaction 
can significantly increase the random motion of stars, making 
a system dynamically hot. Indeed, as shown by some recent 
simulations \citep[see e.g.][]{Sotillo-Ramos2022}, 
newly-born stars during merger tend to have chaotic orbits. 
This might explain why this sub-population is dominated by FSGs. 

Among the interacting FSGs, three galaxies, MaNGA-9182-9102, MaNGA-8937-3703 
and MaNGA-8248-6104, are particularly interesting.
They are identified as FSGs (based on D4000) but have very low star formation rate 
($\log\rm sSFR < -12$). Two of them can be found in the upper-third 
and the lower-second panels of Figure \ref{fig_ssfr-sigc}, respectively. 
The third one has $\log(\sigma_{\rm in}/\sigma_{\rm hot})=0.69$ and thus lies 
outside of the boundary of the figure. These galaxies have small D4000 
over the entire galaxies, implying that they are young, 
but their current specific star formation rates are low. 
This may suggest that their star formation is quenched 
recently by a fast process associated with a recent violent 
interaction. These three galaxies will evolve leftwards in the 
Two-$\sigma$ diagram as they relax to dynamical equilibrium,   
eventually joining the FQG population.

\subsection{Outliers}\label{sec_spe}

In the two lowest stellar mass bins, some FQGs and PQGs 
are located far from the dominating L-shape distribution 
in the sSFR-$\sigma_{\rm in}$ plane (Figure \ref{fig_ssfr-sigc}). 
They have small $\sigma_{\rm in}$ and low sSFR. 
In addition, a small fraction of FQGs have $\sigma_{\rm out}$ 
significantly lower than the main population (Figure \ref{fig_2sig}).
The significance of the presence of these outliers is difficult 
to evaluate. One possibility is that these galaxies are disk-like, 
but they are satellite galaxies or splash back galaxies 
that have been strongly quenched by environmental processes. 
The satellite fraction decreases with increasing stellar mass, and so
this possibility is higher for galaxies of lower mass. 
This is consistent with the fact that these galaxies have relatively 
low stellar mass, as can be seen in Figure \ref{fig_ssfr-sigc}.
However, some of these galaxies are massive, 
with $M_*\sim 10^{11}\Msun$, for which environmental effects 
are expected to be weak \citep[e.g.][]{Peng2010, LiP2020}. Some 
other processes may be responsible for quenching 
such massive and cold galaxies. 


Figure \ref{fig_ms_sigc_QCCs} shows that some QCCs have very low $\sigma_{\rm in}$ 
(see also Figure \ref{fig_2sig}). There are in total 27 
QCCs with $\log{\sigma_{\rm in}}<\log{\sigma_{\rm q}(M_{\rm q})}-2\Delta_{\rm q}$. 
We checked their $\sigma$ profiles and found that 17 of them exhibit a clear 
drop of $\sigma$ in the inner region. The upper-right panel in Figure \ref{fig_2sig} 
shows that a small fraction of PQGs have smaller $\sigma_{\rm in}$ 
than the main population. We also checked their $\sigma$ profiles and found 
that about 60\% of them show a clear drop of $\sigma$ in the inner region. 
We also checked the $\sigma$ profiles of FSGs with 
$\log(\sigma_{\rm out}/\sigma_{\rm hot})>-0.4$ and 
$\log(\sigma_{\rm in}/\sigma_{\rm hot})<-0.5$, and found a
similar drop in the central region. 
This is not observed in normal galaxies, including FQGs, 
where the velocity dispersion almost always declines gradually with radius. 
More investigation is needed to understand the cause of the drop
of $\sigma$ in the inner regions of these galaxies. 

\begin{figure*}
    \centering
    \includegraphics[scale=0.5]{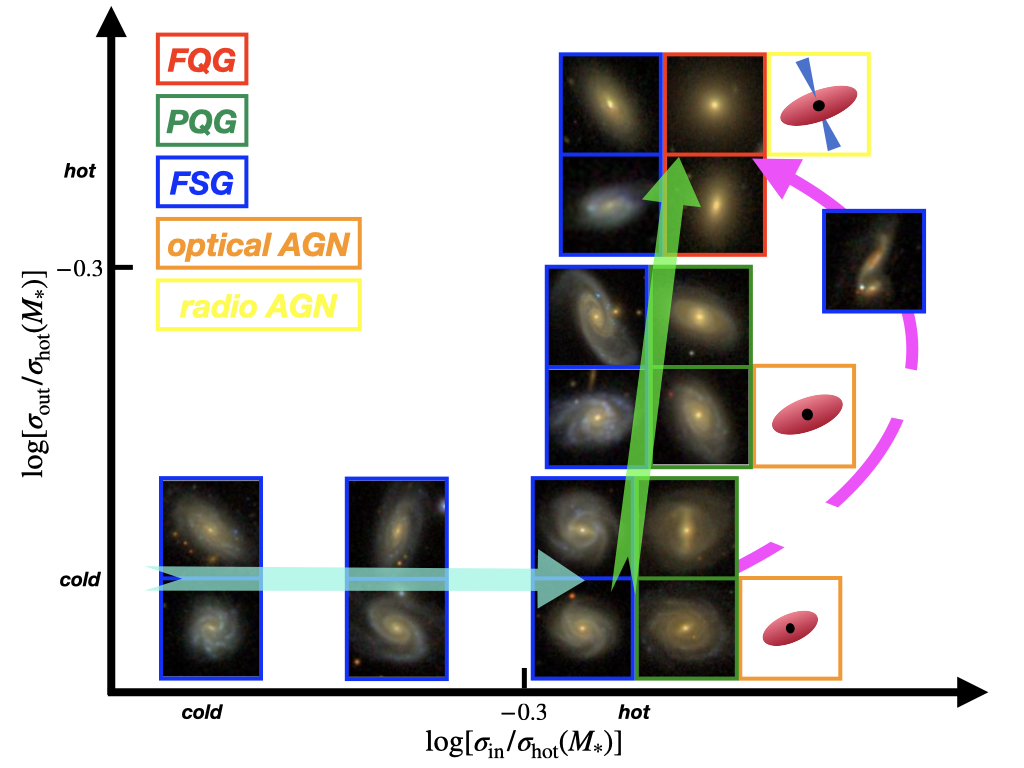}
    \caption{Schematic Two-$\sigma$ diagram. The arrows illustrate possible evolutionary tracks (see text for details).
    We also show the SDSS images of several representative galaxies located at different places along the horizontal and vertical sequences. Red, green and blue square frames indicate the images for FQGs, PQGs and FSGs, respectively.  The orange and yellow frames indicate the sketches for optical and radio AGNs. 
    }\label{fig_evo}
\end{figure*}

\subsection{Evolutionary paths and jumps to dynamical 
hotness and quenching}\label{sec_track}

The Two-$\sigma$, sSFR-$\sigma_{\rm in}$ and sSFR-$\sigma_{\rm out}$ 
diagrams provide information about the co-evolution of galaxy 
star formation and dynamical structure. Here we use the 
Two-$\sigma$ diagram to illustrate these evolutionary paths, 
as shown in Figure \ref{fig_evo}. We emphasize that our discussion 
makes the assumption that the cosmic evolution of these diagrams 
is not large. This assumption has some observational support. 
As shown in \cite{Barro2017ApJ}, the sSFR-$\Sigma_1$ diagram 
is almost the same for galaxies at different redshift. 
However, the evolution of the Two-$\sigma$ diagram is unknown. 
Moreover, the evolutionary paths shown in Figure \ref{fig_evo}, 
and discussed below only represent average trends; 
the paths taken by individual galaxies may be more complicated.

We envisage two major evolutionary paths. The first one is 
represented by the horizontal light-blue arrow, and corresponds to
the horizontal sequence in the Two-$\sigma$ diagram. 
To the left, the galaxy population is dominated by star-forming 
disks, as shown by the representative galaxies in Figure \ref{fig_evo}. 
These disk galaxies may grow by gas accretion and star formation. 
The stellar surface density in the inner region may 
increase because of secular evolution and accretion of gas clouds.  
These galaxies maintain disk morphology and are relatively cold  
in the inner part, as long as the mass of the accreted material 
is insufficient to make the inner part hot.  
This process may lead to increases in the central velocity 
dispersion while the outer part remains dynamically cold. 
Bar and AGN activities may develop, but only at a low level 
in most of the galaxies, because most of the gas component
in the inner part is still supported by angular momentum. 
Star formation can continue, as feedback is too weak to quench star 
formation even in the inner part. A galaxy can continue to evolve in 
this way until its central velocity dispersion reaches about half 
of $\sigma_{\rm hot}(M_*)$. 

A galaxy can become centrally hot when it is significantly 
disturbed, most likely by merger and interaction events that can cause significant 
loss of angular momentum in stars and gas clouds. If only 
minor mergers are involved,  only the central part of a galaxy can 
be affected. The loss of angular momentum allows formed stars 
and gas clouds to sink toward the central region, making the 
galaxy more concentrated. If the merger is gas rich, 
star bursts and AGN activities can be triggered,  
and the associated feedback can start to affect the gas
component and suppress star formation. 
The coexistence of FSGs and PQGs in the lower-right corner of the 
Two-$\sigma$ diagram suggests that 
either AGNs can be triggered only in some, but not all, galaxies,  
or some galaxies can regain cold gas supply in the dynamically hot region   
to offset the feedback. For PQGs where quenching has already occurred 
in the inner part, feedback should be effective only in the dynamically hot part, 
as most of the QCCs are dynamically hot. Physically, this may suggest
that feedback is more effective in a more isotropic distribution 
of gas clouds than in a rotationally supported disk,
as shown in simulations \citep[e.g.][]{Wagner2013}. 
Together, the processes described above produced the first L-shaped 
distribution shown in the left panel of Figure \ref{fig_hs_activity}. 

Centrally hot galaxies in the lower right corner can evolve upwards, 
as indicated by the vertical green arrow in Figure \ref{fig_evo}.
As shown in Figure \ref{fig_vs_morph}, galaxies gradually 
shrink along this sequence, which may be caused by a compaction 
process similar to that discussed in \cite{Barro2017ApJ}.
If the compaction of a galaxy is due to angular momentum loss produced
by interactions and mergers, the orbits of stars and 
star-forming clouds can be randomized, making the galaxy hotter.   
Since the central region is already hot, i.e. already dominated 
by random motion, the effect is expected to be more important in the 
outer region where stars and gas clouds were still in a rotationally 
supported disk. This makes the galaxy move almost vertically 
in the Two-$\sigma$ diagram. However, for a galaxy where 
$\sigma_{\rm in}$ is still significantly lower than 
$\sigma_{\rm hot}(M_*)$, the inner part can be `heated' 
further so that the galaxy can move rightwards within the vertical 
sequence. As long as $\log[\sigma_{\rm out}/\sigma_{\rm hot}(M_*)]<-0.3$, 
most galaxies will remain as FSGs and PQGs. During the process, 
gas clouds that have lost rotational support can sink toward
the center of a galaxy to feed the central black hole.  
AGN feedback may then expel the gas in the hot core, suppressing 
and eventually quenching star formation there. Meanwhile, 
star formation outside the hot core will continue 
as the feedback is not effective in affecting the gas in 
the disk. For some galaxies, quenching may not be important  
as their $\sigma_{\rm out}$ increases and their sizes 
shrink. These galaxies, which remain star forming, although 
dynamically hot, may be observed as blue compact galaxies 
which are as dynamically hot as FQGs 
\citep[see e.g.][]{Barro2017ApJ}. 

When the total effect of galaxy interaction and merging 
is sufficiently strong, a galaxy will become dynamically hot
over its entire body, and moves to the top right corner in the 
Two-$\sigma$ diagram (see the representative galaxies in this region in Figure \ref{fig_evo}). During this process, the gas over the 
entire galaxy can be affected by feedback effects. 
These galaxies will then become fully quenched quickly 
as strong AGN feedback expels gas out of galaxies,
or slowly as star formation consumes cold gas. The fact that 
FQGs and QCCs are all dynamically hot indicates that quenching 
of star formation is effective only in a dynamically hot system.

Massive dynamically hot galaxies are usually surrounded by hot gas halos. 
Accretion of hot gas by the central SMBH can ignite 
`low-excitation' radio AGNs \citep{Best2012}. 
These radio galaxies can heat the circum-galactic medium 
through radio-mode feedback \citep[e.g.][]{Croton2006},    
and prevent gas cooling and further star formation. 
These galaxies can thus become and remain as quenched galaxies. 
Low-mass dynamically hot galaxies may have a different fate: 
it may become totally quenched if no new star-forming gas is 
available or can maintain their star formation if they can acquire
new gas. If the halo is not massive, cooling of the circum-galactic 
medium may occur, adding cold gas in the outskirt and forming a new
disk. In this case, the galaxy may return to the PQG population.  
On the other hand, for a low-mass galaxy living in a dense 
environment, the surrounding gas may be too hot or moving 
too fast for it to capture, and the galaxy will eventually become 
fully quenched. This is consistent with the fact that dynamically hot, 
low-mass galaxies are diverse in their star formation. This
suggests again that being dynamically hot is only a necessary, 
but not sufficient condition for star formation quenching.  

The evolution along the horizontal and vertical sequences 
described above may occur in two different ways. The first   
is through some secular processes, such as the ones driven by 
bars and  minor interactions. As shown in Figure \ref{fig_2sig}
(see also the representative galaxies in \ref{fig_evo}), most of 
the barred galaxies lie along the vertical sequence, 
suggesting that bar-driven processes may be able to re-distribute 
gas and stars, making galaxies centrally hot.  Bar-driven processes 
may also help trigger AGN activity, thereby affecting star formation. 
A recent investigation based on weak lensing and two-point correlation 
analyses \citep{ZhangZ2021} provides some evidence for the role of 
minor interactions in triggering AGNs and building central mass 
concentration. They found that optically selected AGNs and star-forming 
galaxies of the same mass reside in halos of similar mass, but  
that AGNs are surrounded by more satellites. Furthermore, they 
also found that galaxies with larger central velocity dispersion 
tend to have more satellites.  These results suggest that interactions 
with satellites may play an important role not only in 
dynamically `heating' galaxies, but also in triggering AGNs 
and associated feedback. The second way is through violent 
interactions that make jumps along the two sequences in the 
Two-$\sigma$ diagram.  There is evidence for strong 
interactions produced by major mergers in our sample of galaxies
(see Section \ref{sec_vs}). 
Such a violent process can 
significantly enhance dynamical hotness of galaxies and  
make a dynamically cold galaxy completely hot. 
Such interactions can also trigger strong AGNs, and the associated 
feedback may heat and/or expel gas effectively, quenching 
current star formation in galaxies. The associated process 
is expected to be fast and lead to jumps in the Two-$\sigma$ diagram, 
as represented schematically by the pink dashed arrow in Figure \ref{fig_evo}. 



\section{Connection to supermassive black holes}\label{sec_QGBH}

As discussed above, our results indicate that AGN activities 
are related to dynamical hotness of their host galaxies. 
Given that AGN feedback, which has been considered as one of the main
processes of star formation quenching 
\citep[e.g.][]{Silk98, Croton2006, Heckman2014},
is driven by the accretion of supermassive black holes (SMBHs), 
it is interesting to investigate how the growth of SMBH fits in the 
evolutionary scenario proposed above. In this section, we present 
more analyses along this line. In particular, we propose that 
the formation of dynamically hot systems, the growth of SMBHs and 
quenching of star formation are all closely related.  


\begin{figure*}
    \centering
    \includegraphics[scale=0.58]{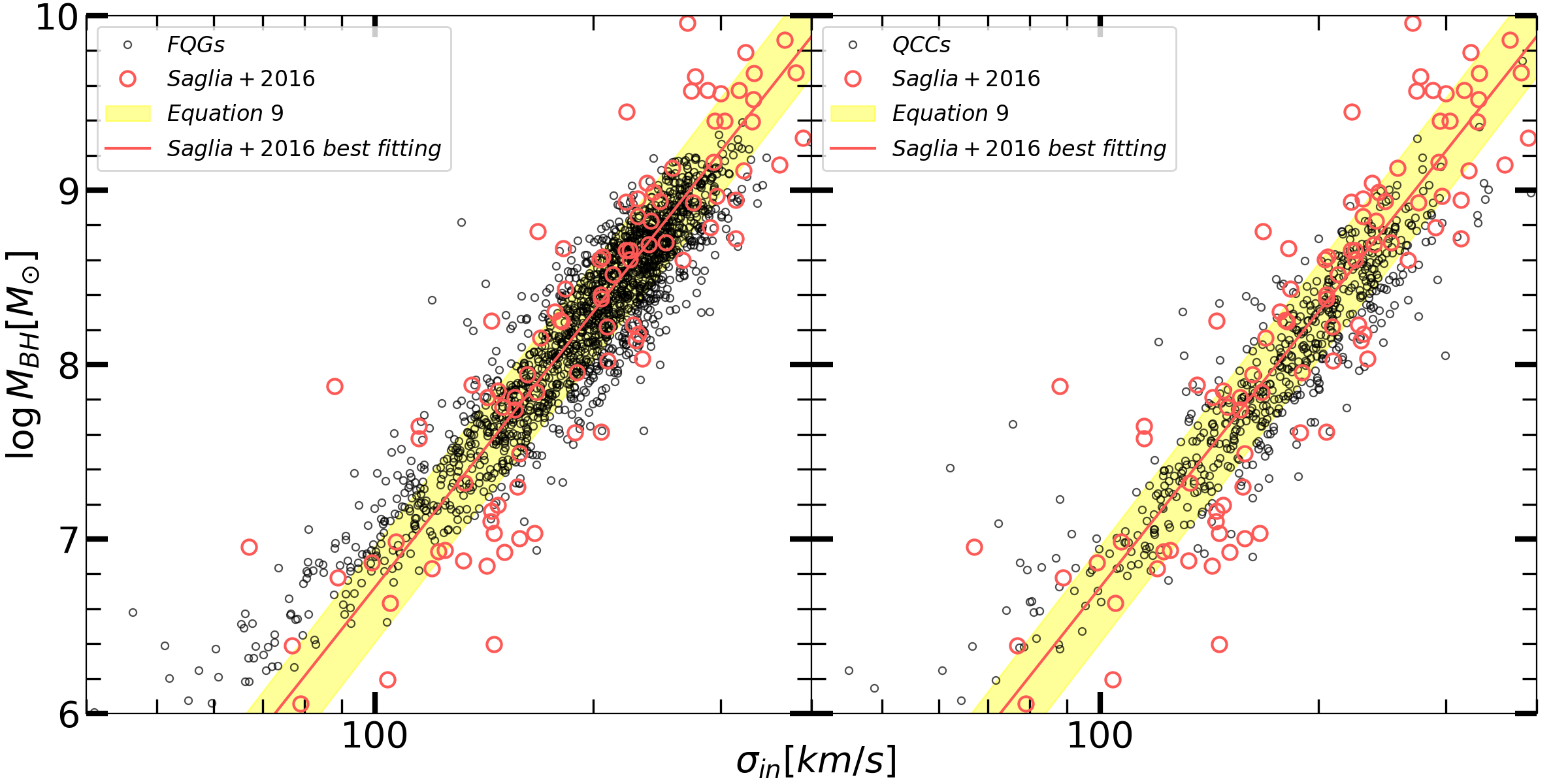}
    \caption{The black circles show the predicted $M_{\rm BH}$ versus $\sigma_{\rm in}$ for FQGs (left) and QCCs (right). The $M_{\rm BH}$ is calculated using Equation \ref{eq_pmbhsig1}. The yellow bands
    show the predicted $M_{\rm BH}$-$\sigma_{\rm in}$ relation (Equation \ref{eq_pmbhsig2}) and its one-$\Delta$ uncertainty based on the best-fitting of the QGSR. The red circles show the observational data (BH1 sample) with dynamically measured black hole masses, taken from \cite{Saglia2016ApJ}. The solid red lines show the
    best-fitting of the BH1 sample, also taken from \cite{Saglia2016ApJ}.
    }\label{fig_mbh-sigc}
\end{figure*}

\begin{figure*}
    \centering
    \includegraphics[scale=0.58]{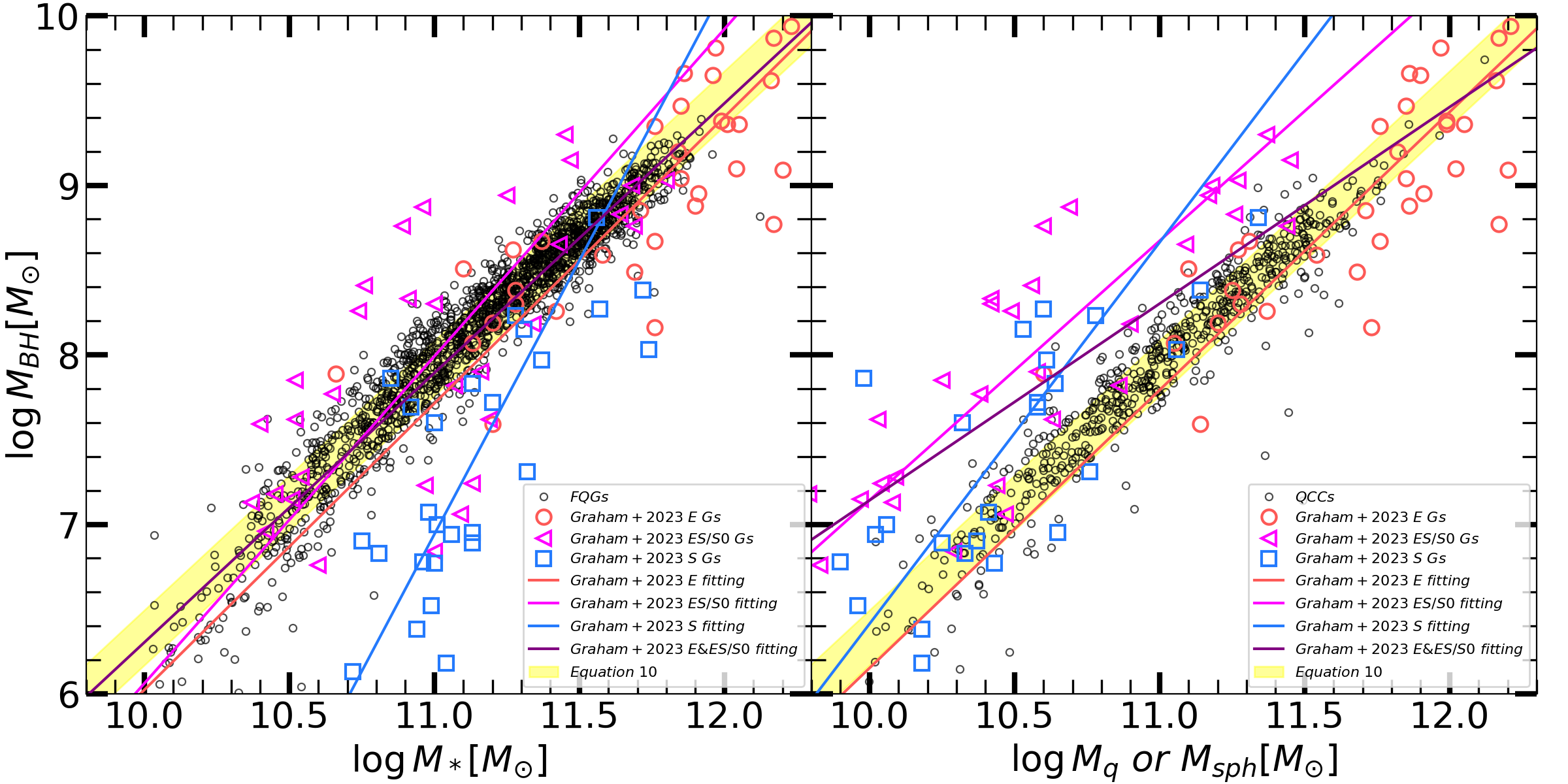}
    \caption{
    The black circles show the predicted $M_{\rm BH}$ versus $M_*$ relation 
    for FQGs (left) and the predicted $M_{\rm BH}$ versus $M_{\rm q}$ relation for QCCs (right). The $M_{\rm BH}$ is calculated using Equation \ref{eq_pmbhsig1}. The yellow bands
    show the predicted $M_{\rm BH}$-$M_*$ relation (Equation \ref{eq_pmbhsig3}) and its one-$\Delta$ uncertainty based on the best-fitting of the QGSR. The color-coded symbols and lines show the observational data (BH2 sample) with dynamically measured black hole masses and their best-fittings, taken from \cite{Graham2023}. The red circles are elliptical galaxies, pink triangles are ES/S0 galaxies, and blue squares are spiral galaxies. 
    The left panel show $M_{\rm BH}$ versus $M_*$ while
    the right panel show $M_{\rm BH}$ versus $M_{\rm sph}$, where $M_{\rm sph}$ is the mass 
    of the spheroid component in a galaxy, derived using the decomposition technique. The purple/red/pink/blue lines show the best fitting for early-type (elliptical$+$ES/S0), 
    elliptical, ES/S0 and spiral galaxies, respectively.
    We convert their Kroupa \citep{Kroupa2002Sci} IMF to our Salpeter \citep{Salpeter1955ApJ} IMF by adding 0.30 dex in the stellar mass (see table 2 in \cite{Bernardi2010MNRA}). 
}\label{fig_mbh-mg}
\end{figure*}

\subsection{A simple model}\label{sec_model}

Let us consider a gas/star system with a given total mass.  
At any given time, the supermassive black hole mass is $M_{\rm BH}$
and the stellar mass is $M_*$. We write the total energy released by the 
accretion of the SMBH as $E=\epsilon M_{\rm BH}c^2/(1-\epsilon)$, where $\epsilon$ 
is the mass-to-energy efficiency and $c$ is the speed of light. 
Only a fraction of the energy is expected to be coupled with 
the inter-stellar medium (ISM) to prevent it from forming stars
either by ejecting the gas or by keeping the gas hot. Denoting this fraction 
as $f_{\rm cp}$, we can write the total gas mass that a SMBH prevents 
from turning into stars as 
\begin{equation}
    M_{\rm gas}=\frac{f_{\rm cp} \epsilon M_{\rm BH}c^2}{1/2 (1-\epsilon) v^2_{\rm esc}}
    =\frac{2\epsilon f_{\rm cp}M_{\rm BH}c^2}{(1-\epsilon)a^2\sigma_{\rm in}^2}\\,
\end{equation}
where $v_{\rm esc}$ is the escape velocity. Since $v_{\rm esc}$ is directly 
related to the gravitational potential of the galaxy, we assume that 
$v_{\rm esc}=a\sigma_{\rm in}$. The SMBH continues to grow its mass until 
most of gas in the host galaxy is affected by the feedback 
and star formation is also quenched. We write the gas-to-star mass ratio 
as  $f_{\rm gas}=M_{\rm gas}/M_*$. The SMBH mass in the quenched object 
(galaxy or QCC) can then be written as
\begin{equation}
    M_{\rm BH}=\frac{(1-\epsilon)a^2f_{\rm gas}M_*\sigma_{\rm in}^2}{2\epsilon f_{\rm cp}c^2}
    =\gamma M_*\sigma_{\rm in}^2
\\,\label{eq_pmbhsig1}
\end{equation}
where $\gamma\equiv (1-\epsilon)a^2f_{\rm gas}/2\epsilon f_{\rm cp}c^2$.

As shown in Section \ref{sec_dhot}, most quenched objects (FQGs and QCCs) 
are dynamically hot. Thus, the above relation is expected to be valid 
only for dynamically hot systems where 
$\sigma_{\rm in} \approx \sigma_{\rm hot} (M_*)$. As one can see, 
if $\gamma$ is roughly a constant, the above relation, combined 
with the $\sigma_{\rm in}$-$M_*$ relation (i.e. quenched galaxy scaling relation), leads to a relation 
between $M_{\rm BH}$ and $\sigma_{\rm in}$, or between  
$M_{\rm BH}$ and $M_*$. This suggests that we can use the observed 
scaling relations of $M_{\rm BH}$ with $M_*$ and $\sigma_{\rm in}$
to constrain $\gamma$. For this purpose, we combine the 
QGSR given by Equation \ref{eq_qgsr} with Equation \ref{eq_pmbhsig1} to 
obtain 
\begin{equation}
    M_{\rm BH}=\gamma10^{-\beta
/\alpha}\sigma_{\rm in}^{2+1/\alpha}=\gamma10^{-\beta
/\alpha}\sigma_{\rm in}^{\alpha_{\rm MS}}\\,\label{eq_pmbhsig2}
\end{equation}
and
\begin{equation}
    M_{\rm BH}=\gamma10^{2\beta}M_*^{1+2\alpha}
    =\gamma10^{2\beta}M_*^{\alpha_{\rm MM}}\\, \label{eq_pmbhsig3}
\end{equation}
where $\alpha=0.30$ and $\beta=-1.02$, as obtained in Equation \ref{eq_qgsr}
(see Table \ref{tab_term}). As a simple model, we assume that $\gamma$ is a constant. 
In this case, we have only one model parameter, and we use the data 
of $M_{\rm BH}$ and velocity dispersion presented in \cite{Saglia2016ApJ} 
(hereafter the BH1 sample) for this purpose. The sample consists of 96 
galaxies and we use all of them to obtain a best estimate for $\gamma$. 
This gives $\gamma= 2.87\times 10^{-8}$, which completely specifies the 
two scaling relations given by  Equations \ref{eq_pmbhsig2} and 
\ref{eq_pmbhsig3}.  


\subsection{The validity of the model}\label{sec_bhsr}

With the value of $\gamma$ calibrated, Equation \ref{eq_pmbhsig1} can 
be used to predict $M_{\rm BH}$ for each FQG and QCC. 
For QCCs, we use the mass within $R_{\rm q}$, i.e. 
$M_{\rm q}$ (see Section \ref{sec_gclass}) as $M_*$. 
The predicted $M_{\rm BH}$ versus $\sigma_{\rm in}$ relations
for FQGs and QCCs are shown, respectively, as black circles in the 
left and right panels of Figure \ref{fig_mbh-sigc},  
together with Equation \ref{eq_pmbhsig2} 
(the yellow band in each panel).
For comparison, the $M_{\rm BH}$ versus $\sigma$ for galaxies in the 
BH1 sample \citep[][]{Saglia2016ApJ} are shown as red circles together  
with its best-fitting to a power-law relation between $M_{\rm BH}$ and 
$\sigma_{\rm in}$ (red lines). As one can see, the predicted $M_{\rm BH}$ 
is tightly correlated with $\sigma_{\rm in}$ for both FQGs and QCCs, and 
both populations follow the same trend described by Equation \ref{eq_pmbhsig2}. 
Our model prediction also follows closely the trend defined by the 
BH1 sample. This is not trivial, as the host galaxies of QCCs are PQGs,
which are morphologically different from FQGs, and consistent with 
the assertion that galaxies of different morphological types follow 
a similar $M_{\rm BH}$-$\sigma_{\rm in}$ relation \citep[e.g.][]{Kormendy2013ARA&A}. 
The slope of our $M_{\rm BH}$-$\sigma_{\rm in}$ relation can be 
obtained directly from the QGSR. Using Equation \ref{eq_pmbhsig2}, 
we obtain $\alpha_{\rm MS}=2+1/\alpha=5.39$, in good agreement
with the slope, $5.25\pm0.27$, obtained from the BH1 sample 
\citep{Saglia2016ApJ}. 

Note that \cite{Saglia2016ApJ} adopted the effective velocity dispersion, 
$\sigma_{\rm e}$, defined as the luminosity-weighted mean value of 
$\sqrt{V^2(r)+\sigma^2(r)}$ within $1 R_{\rm e}$, with 
$V(r)$ and $\sigma (r)$ being the line-of-sight velocity and velocity dispersion, respectively.
This is different from our $\sigma_{\rm in}$. Our test shows that the 
$\sigma_{\rm in}$-$\sigma_{\rm e}$ relations for FQGs and QCCs have very small scatter 
and lie closely to the 1:1 relation. It is consistent with the fact that these objects 
are hot systems, i.e. dispersion-dominated, and their dispersion profiles are flat. 
In fact, \cite{Saglia2016ApJ} also found that the dispersion is almost 
independent of radius. We thus anticipate that our results remain unchanged 
even if $\sigma_{\rm e}$ is used in Figure \ref{fig_mbh-sigc}.


In Figure \ref{fig_mbh-mg}, we present the predicted $M_{\rm BH}$-$M_*$ ($M_{\rm q}$) 
relation for FQGs (QCCs), together with Equation \ref{eq_pmbhsig3} for reference. 
The two masses are tightly correlated, and both FQGs and QCCs follow the 
a similar relation that matches Equation \ref{eq_pmbhsig3}. 
\cite{Saglia2016ApJ} did not provide stellar mass for galaxies in the BH1 sample. 
We thus compare our prediction with the black hole sample 
of \cite{Graham2023} (hereafter the BH2 sample) that provides 
stellar mass of host galaxies. Interestingly, our FQGs follow closely 
early-type galaxies (ellipticals and ES/S0; left panel of Figure \ref{fig_mbh-mg})
in BH2. \cite{Graham2023} performed a linear regression 
for early-type galaxies and found a slope of $1.59\pm0.11$, which is shown 
as the purple line in the left panel. The slope of our prediction is 
$\alpha_{\rm MM}=1+2\alpha=1.59$ (Equation \ref{eq_pmbhsig3}).
\cite{Graham2023} also provided fitting results for ellipticals and ES/S0, 
separately (the red and pink lines in Figure \ref{fig_mbh-mg}). 
As one can see, the two lines are consistent with our prediction 
at the high- and low-mass ends, respectively. 
The discrepancy between the  BH2 spiral galaxies (blue squares and line) 
and our FQGs is very large, as is expected because spiral galaxies are 
usually star-forming and our model only applies to quenched objects.

\cite{Graham2023} also provided the mass of the spheroid component
($M_{\rm sph}$) for each galaxy. The spheroid component represents an elliptical 
or the bulge of a spiral. It is thus interesting to compare results of 
our QCCs with those of spheroids, and the comparison is shown 
in the right panel of Figure \ref{fig_mbh-mg}. Our QCCs follow closely 
elliptical galaxies (red circles and line) in the BH2 sample. However, 
the results for bulges of ES/S0 galaxies (pink triangles and line) are very different.  
\cite{Graham2023} also found a significant offset in the 
$M_{\rm BH}$-$M_{\rm sph}$ relation between ellipticals and bulges
(see also Figure \ref{fig_mbh-mg}). At given $M_{\rm BH}$, the bulge mass 
is about 0.5 dex smaller than that of the QCCs.  
Recently, \cite{Zhu2022b} decomposed stellar components of two nearby 
early-type galaxies using dynamical information obtained from 
IFU spectroscopy. In addition to bugles, they found another dynamically 
hot component within $2R_{\rm e}$ of the galaxies, and referred to it  
as a dynamically hot inner stellar halo (DHISH). The mass of DHISH is about 
30\% of the total stellar mass. Similar components can also be found in 
simulated galaxies \citep{Zhu2022a}, with the mass fraction ranging from 
several to more than fifty percents. The tension between our prediction 
and the observed $M_{\rm BH}$-$M_{\rm sph}$ relation can 
thus be alleviated by including DHISH. 

To quantify the performance of our model, we adopt two $\chi^2$ parameters 
to measure the differences between the model prediction and observational data.
For BH1 the $\chi^2$ is:
\begin{equation}
    \chi_1^{2}=\frac{1}{N_1}\sum_{i=1}^{N_1} \frac{(\log M_{\rm BH}(\sigma_{i})-\log M_{{\rm BH1},i})^2}{e^2_{M_{{\rm BH1},i}}+\alpha^2_{\rm MS}e^2_{\sigma_{i}}+\frac{1}{\alpha^2}\Delta^2_{\rm q}}\\,\label{eq_chi1}
\end{equation}
where $M_{{\rm BH1},i}$ and $e_{M_{\rm BH1},i}$
are the measurement of $M_{\rm BH}$ and its uncertainty
for the $i$th galaxy in BH1, $\sigma_i$ and $e_{\sigma_i}$ are 
the measurement of the velocity dispersion and its uncertainty. 
The data are taken from \cite{Saglia2016ApJ}, and the sample size is $N_1=96$. 
The mass $M_{\rm BH}(\sigma_i)$ in the above equation is the
prediction of Equation \ref{eq_pmbhsig2} for the $i$th galaxy. 
For BH2, the $\chi^2$ is defined as 
\begin{equation}
    \chi_2^{2}=\frac{1}{N_2}\sum_{i=1}^{N_2} \frac{(\log M_{\rm BH}(M_{*,i})-\log M_{{\rm BH2},i})^2}{e^2_{M_{{\rm BH2},i}}+\alpha^2_{\rm MM}e^2_{M_{*,i}}+4\Delta^2_{\rm q}}\,,\label{eq_chi2}
\end{equation}
which is the same as Equation \ref{eq_chi1}, except that it is 
for the $N_2=67$ elliptical and ES/S0 galaxies in the BH2 sample, 
with data taken from \cite{Graham2023}. The black hole mass,
$M_{\rm BH}(M_{*,i})$, is predicted by Equation \ref{eq_pmbhsig3}.  
Note that any covariance in the measurements are ignored, 
but the intrinsic scatter of the QGSR is included. 
We obtain $\chi_1^2=1.80$ and $\chi_2^2=2.17$, indicating 
that our model predictions are consistent with the observational data.



\subsection{Implications of the model}

The fact that our simple model based on scaling relations 
of FQGs and QCCs can reproduce the observed black hole mass 
suggests that quenching of star formation and the growth 
of SMBHs are closely linked, and both are related to the 
dynamically hot stellar component. In our simple model, the parameter
$\gamma$ is assumed to be constant, and its value is well constrained 
by observational data. Since $\gamma$ is a combination of processes 
specified by $a$, $f_{\rm gas}$, $\epsilon$ and $f_{\rm cp}$ (Equation \ref{eq_pmbhsig1}), 
we may use the constrained value to examine its implications 
for the related processes. The parameter $a$ is related to the shape 
and dynamical status of a galaxy. For a virialized, uniform sphere, 
we have $a^2\approx 6$.  The value of $f_{\rm cp}$ is uncertain. 
For an AGN in the high-accretion phase where radiation is the dominating 
feedback channel, $f_{\rm cp}$ is usually assumed to be about $0.05$
\citep[e.g.][]{Springel2005, Hopkins2006}. 
However, the value could be much lower according to the observational 
studies \citep{He2019NatAs, He2022SciA}. 
Assuming that $\epsilon=0.1$ and $a^2=6$, we 
obtain $f_{\rm gas}= 4.8 (f_{\rm cp}/0.05)$. 
Thus, the average amount of gas that can be prevented from forming stars can 
be significant. Note that $f_{\rm gas}$ may vary from one system to 
another. At a given $M_*$,  a black hole in a gas-rich system has to release 
more energy to quench the system than that in a gas-poor system, 
which leads to more growth in the mass of the
black hole. Therefore, the final $M_{\rm BH}$ is expected to depend 
on the gas richness before quenching. This suggests that the intrinsic 
scatter in BHSRs should be larger than the prediction of our 
simple model that does not include the scatters in model parameters.
It is consistent with that both $\chi^2_1$ and $\chi^2_2$ are larger than one.

At low $\sigma_{\rm in}$ end, the predicted $M_{\rm BH}$-$\sigma_{\rm in}$ 
relation systematically deviates from the observational results. 
These black holes usually reside in low-mass galaxies.
As shown in Figure \ref{fig_sigma_prof}, low-mass FQGs usually have larger gradients in their $\sigma$ profiles. The gas in the outskirts
can thus escape more easily, and the value of $a$ is lower, than 
the gas in the outskirts of more massive galaxies. This can  
lower the predicted $M_{\rm BH}$ for low-mass galaxies. 
Furthermore, stellar feedback may play a more important role 
in low-mass galaxies, which can reduce the accretion rate of black holes
and their mass growth, as suggested by \cite{Hopkins2022MNRAS}.

The difference between our QCCs and bugles (Figure \ref{fig_mbh-mg})
suggests that the galaxy-black hole relation defined  
by stellar components is related to the growth of the central 
black hole. \cite{Zhu2022a} analyzed the DHISH components in simulated 
galaxies and found that they are directly produced by galaxy mergers.
These mergers are expected to trigger strong AGNs and cause significant
growth in the mass of the SMBH. This suggests that the DHISH should be an 
important part of the galaxy-black hole relation, and is needed 
to explain the difference in the $M_{\rm sph}$-$M_{\rm BH}$ relation between 
elliptical galaxies and bulges found by \cite{Graham2023}.


\section{Summary}\label{sec_sum}

In this paper, we use MaNGA galaxies to investigate the relationships 
among galaxy dynamical status, galaxy quenching, supermassive black hole mass, 
and the co-evolution of galaxy dynamical and quenching properties. 
The spatially resolved dynamical property, as
represented by stellar velocity dispersion, and stellar population property represented 
by D4000 make it possible to establish the connection in such a way as to 
make physical interpretations more straightforward. 
We only use central galaxies to avoid contamination by environmental effects.
Our main results are summarized below.

Different from most of previous studies, we use the fraction of quenched spaxels
to classify galaxies into three classes, fully quenched (FQGs), partially quenched(PQGs) 
and fully star-forming galaxies(FSGs). We find that most FQGs have very weak star 
formation activity, with $\log\rm sSFR<-12$, while FSGs have strong star formation, 
with $\log\rm sSFR>-10.6$. PQGs have quenched fractions between 5\% and 95\% 
and $\log\rm sSFR$ between those of FSGs and FQGs (Figure \ref{fig_ssfr-sigc}). 
We also identify quenched central cores (QCCs) that have quenched fraction similar to FQGs. 

We define a scaling relation of dynamical hotness (SRDH) using the 
$M_{*}$-$\sigma_{\rm hot}$ relation of dynamically hot galaxies 
(Equation \ref{eq_fjr}). We use the difference between the observed 
velocity dispersion of a galaxy and $\sigma_{\rm hot}(M_*)$ given 
by the SRDH for its stellar mass to describe its dynamical hotness. 
The Two-$\sigma$ diagram, which plots the scaled inner velocity 
dispersion ($\log(\sigma_{\rm in}/\sigma_{\rm hot})$) versus the 
scaled velocity dispersion in the outer part 
($\log(\sigma_{\rm out}/\sigma_{\rm hot})$), Figure \ref{fig_2sig}), 
is used to distinguish between centrally and totally hot galaxies.  
We find that both FQGs and QCCs occupy a small region around 
(0,-0.05) in this diagram, demonstrating that FQGs and QCCs  
are dynamically hot over their entire bodies. In contrast, 
most PQGs and FSGs are located blow the SRDH (Figure \ref{fig_sigc-ms}), 
indicating that they are dynamically colder.
PQGs have a very broad distribution in 
$\log(\sigma_{\rm out}/\sigma_{\rm hot})$ and form an extended 
vertical band in the Two-$\sigma$ diagram (Figure \ref{fig_2sig}).
This suggests that PQGs are dynamically hot in central regions 
but dynamically cold in outskirts. FSGs have an even broader 
distribution in dynamical state: they can not only reside in the 
vertical band defined by PQGs, but also in a horizontal sequence 
where a galaxy is cold both in the inner and outer regions. 

The distribution of galaxies in the Two-$\sigma$ diagram
shows clear evolutionary connections and transitions 
of galaxy properties, such as morphology, star formation and 
AGN activity (Figure \ref{fig_evo}). We define a horizontal sequence 
($\log(\sigma_{\rm out}/\sigma_{\rm hot})<-0.5$), in which 
galaxies remain dynamically cold in outer regions,  
and a vertical sequence ($\log(\sigma_{\rm in}/\sigma_{\rm hot})>-0.3$), 
in which galaxies are centrally hot (Figure \ref{fig_2sig}), 
to follow the evolution of the galaxy population. 
On the horizontal sequence, centrally cold galaxies  
are dominated by star-forming disks, and grow via star formation. 
The central velocity dispersion increases gradually 
while the outskirt remains cold. AGN activities are rare 
and so no significant quenching of star formation occurs.
When the central dispersion of a galaxy reaches about half of the value 
given by the SRDH according to its mass, the galaxy 
becomes more concentrated and morphologically more diverse.
Optical AGNs then appear frequently and galaxies 
can be quenched in central regions, leading to the emergence of 
QCCs and PQGs. As the outskirts of cold disks are heated up 
gradually, galaxies move upwards along the vertical sequence,
eventually reaching the top of the vertical sequence and 
becoming dynamically hot. Their star formation can be 
quenched if strong AGNs are triggered during the process.
For a hot massive galaxy, accretion of hot gas may ignite 
radio AGN, which can heat the surrounding gas and keep 
the galaxy quenched.  All these processes seem to lead 
to two transitions in the sSFR-$\sigma_{\rm in}$ and 
sSFR-$\sigma_{\rm out}$ diagrams when galaxies move along 
the horizontal and vertical sequences
(Figure \ref{fig_hs_activity} and \ref{fig_vs_activity}). 

Our results suggest that both secular and rapid processes can drive
the evolution and change of galaxy dynamical status (Figure \ref{fig_2sig}). 
We find that barred galaxies tend to lie exactly on the vertical sequence, 
consistent with the fact that bars can redistribute gas and stars 
and drive the growth of central concentration. Galaxies with large
central velocity dispersion and dynamically hot FSGs both tend to 
exhibit signatures of strong interactions and mergers. This indicates 
that these processes can change the dynamical state of a galaxy 
very quickly. We find that optically selected AGNs tend to lie on 
the vertical sequence where quenching starts and that radio galaxies 
can only be found in dynamically hot galaxies where the evolution 
in the dynamical state of a galaxy cannot proceed further. 
These results suggest that AGNs are connected to the change of 
the dynamical status and quenching of galaxies,  
consistent with the hypothesis that AGN feedback plays 
an important role in quenching star formation.

We construct a simple model to link the energy released by AGNs 
and the energy required to get rid of star-forming gas. 
Since our observational results show that only dynamically hot systems 
can be quenched, the model applies only to quenched and dynamically 
hot systems, such as FQGs and QCCs. We use the model  
to predict the mass of SMBHs, $M_{\rm BH}$, and find 
the predicted $M_{\rm BH}$-$\sigma_{\rm in}$ and 
$M_{\rm BH}$-$M_*$ relations are in good agreement with observational
data (Figure \ref{fig_mbh-sigc} and \ref{fig_mbh-mg}). 
All these show that the formation of dynamically hot systems, 
the growth of SMBHs and quenching of star formation are closely 
connected with each other.




\section*{Acknowledgements}
We thank Kai Wang and Huiling Liu for helpful discussions. We thank the referee for useful suggestions. This work is supported by the National Key R\&D Program of China (grant No. 2018YFA0404503), the National Natural Science Foundation of China (NSFC, Nos. 12192224, 11733004 and 11890693), CAS Project for Young Scientists in Basic Research, Grant No. YSBR-062, and the Fundamental Research Funds for the Central Universities. We acknowledge the science research grants from the China Manned Space Project with No.
CMS-CSST-2021-A03. The authors gratefully acknowledge the support of Cyrus Chun Ying Tang Foundations. The work is also supported by the Supercomputer Center of University of Science and Technology of China.

\bibliography{ref.bib}


\clearpage
\end{document}